\documentclass[aps, prd, 10pt, twocolumn, superscriptaddress,noshowpacs, preprintnumbers, longbibliography,nofootinbib,bibnotes,hyperref,floatfix]{revtex4-1}

\usepackage{amsmath}
\usepackage{amsfonts}
\usepackage{amssymb}
\usepackage{bbold}
\usepackage{epsfig}
\usepackage{graphicx}
\usepackage{bm}
\usepackage{array}
\usepackage{hyperref}
\usepackage{listings}
\usepackage{color}
\usepackage{float}
\usepackage{url} 
\usepackage[normalem]{ulem}

\usepackage{lineno}
\usepackage{lipsum}
\usepackage{physics}
\usepackage{mathtools}

\usepackage{tabularx}
\usepackage[export]{adjustbox}
\usepackage{multirow}
\usepackage{citesort}
\usepackage{soul}
\usepackage{bm}
 \usepackage{xcolor}
\usepackage[utf8]{inputenc}
\usepackage[normalem]{ulem}
\usepackage{hyperref}
\usepackage{accents}
\usepackage{tikz}

\usepackage{tikz,xcolor,hyperref}

\definecolor{lime}{HTML}{A6CE39}
\DeclareRobustCommand{\orcidicon}{\hspace{-1mm}
	\begin{tikzpicture}
	\draw[lime, fill=lime] (0,0) 
	circle [radius=0.16] 
	node[white] {{\fontfamily{qag}\selectfont \tiny \,ID}};
	\draw[white, fill=white] (-0.0525,0.095) 
	circle [radius=0.007];
	\end{tikzpicture}
	\hspace{-3mm}
}

\foreach \x in {A, ..., Z}{\expandafter\xdef\csname orcid\x\endcsname{\noexpand\href{https://orcid.org/\csname orcidauthor\x\endcsname}
			{\noexpand\orcidicon}}
}

\begin{document}

\title{Energy Dependence of Flavor Instabilities Stemming from Crossings in the  Neutrino Flavor Lepton Number Angular Distribution}

\author{Pedro Dedin Neto\orcidA{}}
\affiliation{Niels Bohr International Academy \& DARK, Niels Bohr Institute, University of Copenhagen, Blegdamsvej 17, 2100 Copenhagen, Denmark}
\affiliation{Universidade Estadual de Campinas, Instituto de F\'{\i}sica Gleb Wataghin, R. S\'ergio Buarque de Holanda, 777, Brazil}
\author{Irene Tamborra\orcidB{}}
\affiliation{Niels Bohr International Academy \& DARK, Niels Bohr Institute, University of Copenhagen, Blegdamsvej 17, 2100 Copenhagen, Denmark}
\author{Shashank Shalgar\orcidC{}}
\affiliation{Niels Bohr International Academy \& DARK, Niels Bohr Institute, University of Copenhagen, Blegdamsvej 17, 2100 Copenhagen, Denmark}


\date{\today}

\begin{abstract}
In core-collapse supernovae and neutron star mergers, the neutrino density is so large that neutrino-neutrino refraction can lead to flavor conversion, if a zero-crossing is present in the neutrino flavor lepton number (FLN) angular distribution and
the neutrino self-interaction strength 
$\mu=\sqrt{2} G_F n_\nu$ represents the characteristic timescale of the system. It has been empirically realized that the vacuum frequency $\omega=\Delta m^2/2E$ affects the development of flavor conversion in the presence of zero-crossing even if $\omega \ll \mu$. Focusing on a homogeneous and axially symmetric neutrino gas, we explore the role of $\omega$ in the onset of flavor instabilities.
We find that a non-zero vacuum frequency can be responsible for inducing flavor instabilities even when the neutrino self-interaction strength  is much larger than the vacuum frequency. Moreover, mapping a neutrino ensemble with $\omega \neq 0$  into an effective system with $\omega =0$, we find that a system with no FLN zero-crossing can effectively develop one for $\omega \neq 0$ becoming unstable. 

\end{abstract}

\maketitle

\section{Introduction}
\label{sec:Introduction}

In dense astrophysical environments, neutrinos experience refractive effects due to other neutrinos with major implications on their flavor evolution~\cite{pantaleone1992dirac,pantaleone1992neutrino}. Because of this, the equations of motion of neutrinos acquire a nonlinear nature. The rich phenomenology of neutrino-neutrino refraction has been explored for a long time in the context of  ``slow flavor conversion.'' 
 This type of flavor conversion has been dubbed slow, since the vacuum frequency $\omega=\Delta m^2/2E$ (with $\Delta m^2$ being the mass-squared difference and $E$ the neutrino energy) determines the characteristic flavor conversion scale together with the neutrino self-interaction strength  $\mu=\sqrt{2} G_F n_\nu$ (with $G_F$ being the Fermi constant and $n_\nu$ the local density of neutrinos)~\cite{Duan:2006an,Duan:2010bg,Mirizzi:2015eza}. The equations of motion of neutrinos, in this context, formally resemble the ones of a gyroscopic pendulum~\cite{Hannestad:2006nj,Duan:2007mv,Raffelt:2011yb,Fogli:2007bk,Fogli:2008pt,Johns:2017oky}. 
More recently, it has been realized that flavor conversion can occur at higher rates, even in the absence of vacuum mixing (``fast flavor conversion'')~\cite{Sawyer:2005jk,Sawyer:2008zs}. In this case, the only characteristic scale entering the neutrino equations of motion is the neutrino self-interaction energy $\mu$~\cite{Chakraborty:2016yeg,Tamborra:2020cul,Richers:2022zug}. 

Fast flavor conversion is intrinsically a multi-angle phenomenon. A necessary condition for the existence of fast flavor instabilities is the existence of a zero-crossing in the angular distribution of the electron lepton number (ELN) of neutrinos~\cite{Chakraborty:2016lct,Izaguirre:2016gsx,Morinaga:2021vmc}. 
As with the case of slow flavor conversion, for a homogeneous and axially symmetric neutrino system, one can show that the neutrino equations of motion are formally equivalent to the ones of a gyroscopic pendulum~\cite{Padilla-Gay:2021haz,Johns:2019izj,Fiorillo:2023hlk,Fiorillo:2023mze}, where the depth of flavor conversion is linked to the real part of the eigenfrequency resulting from the linear normal-mode analysis~\cite{Padilla-Gay:2021haz}.

While fast flavor conversion can develop in the limit $\mu \gg \omega$, if an ELN angular zero-crossing is present,  the flavor phenomenology can be altered by the vacuum frequency~\cite{Chakraborty:2019wxe,Airen:2018nvp,Zaizen:2021wwl,Shalgar:2020xns,Shalgar:2022rjj,Shalgar:2021wlj}. 
We show that it is possible to trigger a flavor instability in the absence of ELN crossing for large ratios of $\mu/\omega \sim \mathcal{O}(10^{5})$, according to the shape of the angular distribution. Even though such a flavor instability is induced by the vacuum term, it does not abide by the conventional understanding of the slow flavor instability, which is expected to be suppressed at very large densities~\cite{Esteban-Pretel:2008ovd,Esteban-Pretel:2007jwl,Chakraborty:2011nf}.

 The paper is organized as follows. In Sec.~\ref{sec:Evolution_Equation}, we introduce the neutrino equations of motion, the linear normal mode analysis, and the parametrization adopted for the neutrino angular distributions. In Sec.~\ref{sec:Vanishing_w}, we revisit the fast stability conditions for a neutrino system with vanishing vacuum mixing. The impact of non-vanishing vacuum mixing on the development of a flavor instability otherwise triggered by zero-crossing in the ELN  angular distribution is explored in Sec.~\ref{sec:Non_Vanishing_w}; there, first we investigate the role of $\omega\neq 0$ on the development of flavor instabilities through the linear stability analysis. Then, we consider a perturbative approach for neutrino ensembles with and without non-electron flavors. We also investigate the possible mapping of a system with $\omega\neq0$ to an effective one with $\omega=0$ with effective angular distributions, showing that ELN angular distributions without zero-crossing can effectively develop one because of vacuum mixing. Finally, we summarize our findings in Sec.~\ref{sec:Conclusions}. Details on the impact of the matter potential on the development of flavor instabilities stemming from ELN angular zero-crossings are provided in Appendix~\ref{sec:matter}. In Appendix~\ref{sec:Appendix_Numerical}, we outline the method employed for computing the eigenfrequency in the linear stability analysis.

\section{Neutrino equations of motion}
\label{sec:Evolution_Equation}
The flavor evolution of neutrinos and antineutrinos can be investigated relying on the density matrices, $\rho(t, \vec{p}, \vec{x})$ and $\bar\rho(t, \vec{p}, \vec{x})$. The diagonal entries represent the occupation numbers of a given flavor state and the off-diagonal ones encode information on flavor coherence. For the sake of simplicity, we work in the two-flavor approximation ($\nu_e$, $\nu_x$) and consider one energy mode $E$. Even though the system is mono-energetic, we explore the impact of $\omega$ for different energy modes in the following sections. The equations of motion assume the following form in the absence of external forces and collisions:
\label{eq:QKE_rho}
 \begin{equation}
 i\left( \partial_t +\vec{v}\cdot\vec\nabla_{\vec x} \right) \rho(t, \vec p,\vec x) = [H,\rho(t, \vec p,\vec x)]\ ,
 \end{equation}
 \begin{equation}
 \label{eq:QKE_rho1}
 i\left( \partial_t +\vec{v}\cdot\vec\nabla_{\vec x} \right) \bar\rho(t, \vec p,\vec x) = [\bar H,\bar\rho(t, \vec p,\vec x)]\ .
 \end{equation}
The Hamiltonian $H$ is given by the sum of the vacuum term ($H_{\text{vac}}$), the matter term ($H_{\text{matt}}$), and the neutrino-neutrino one ($H_{\nu\nu}$): 
\begin{eqnarray}
 H_{\text{vac}} &=& \frac{\omega}{2} \begin{pmatrix}
 -\cos{2\theta_V} &\sin{2\theta_V} \\ 
 \sin{2\theta_V} & \cos{2\theta_V} 
 \end{pmatrix} \equiv \frac{1}{2}\begin{pmatrix}
 -\omega^c &\omega^s \\ 
 \omega^s & \omega^c 
 \end{pmatrix}\ ,\\
 H_{\text{matt}} &=& \begin{pmatrix}\sqrt{2}G_F n_e&0 \\ 0 &0 \end{pmatrix} \equiv \begin{pmatrix}\lambda &0 \\ 0 &0 \end{pmatrix}\ ,\\
 H_{\nu\nu} &=& \mu \int \mathrm{d}\vec{p'} (1-\hat p \cdot \hat{p'})[\rho(\vec{p'}) - \bar\rho(\vec{p'})]\ .
\end{eqnarray}
Here $\theta_V$ is the vacuum mixing angle, $\omega^c = \omega \cos 2\theta_V$, $\omega^s = \omega \sin 2\theta_V$, and  $n_e$ stands for the local number density of electrons. For antineutrinos, $\bar{H}$ is identical to $H$ except for the fact that $\bar{H}_{\text{vac}}$ is replaced by $- H_{\text{vac}}$~\cite{sigl1993general}.

For the sake of simplicity, we consider a homogeneous neutrino gas with axial symmetry in momentum space. In this approximation, the density matrices take the form:
\begin{equation}
 \rho(t,\vec{p},\vec{x}) \equiv\rho(t,v) ,\;\;\;\;\;\bar\rho(t,\vec{p},\vec{x})\equiv \bar\rho(t,v) \ ,
\end{equation}
where $v\equiv \cos\theta$ represents the projection of the velocity $\vec v = \vec p/E$ along the axis of symmetry, for a neutrino traveling at the speed of light.  
The neutrino-neutrino component of the Hamiltonian, $H_{\nu\nu}$, is
\begin{equation}
\label{eq:H_nunu_normalized}
 H_{\nu\nu} = \mu \int^{+1}_{-1} \mathrm{d}v' (1-v v') D_{v'}\ ,
\end{equation}
where  $D_v\equiv \rho(v)-\bar\rho(v)$.
One can also decompose the Hermitian  matrix $D_v$ through the Bloch vectors: 
\begin{equation}
 D_v = \frac{1}{2} \left(\mathrm{Tr}D_v \mathbb{I}_{2\times2} + \vec D_v \cdot \vec\sigma\right)\ ,
\end{equation}
with $\vec \sigma$ being the vector of Pauli matrices and $\vec D_v$ being a three-dimensional vector whose higher order multipoles are defined as $\vec D_n = \int_{-1}^{+1}  \mathrm{d}v {v}^n \vec D_{v}$.

\subsection{Linear normal mode analysis}
\label{sec:Linear_Stability}
The linear normal mode analysis has been largely employed in the context of neutrino-neutrino interactions in order to predict the existence of flavor instabilities and their growth rate~\cite{Banerjee:2011fj,Airen:2018nvp,Izaguirre:2016gsx,Capozzi:2019lso,Morinaga:2018aug}.
After writing the density matrices in the flavor basis as 
\begin{equation}
\rho (t,v) = \begin{pmatrix}
 \rho_{ee}(t,v)& \rho_{ex}(t,v)\\ 
 \rho_{ex}^*(t,v)&\rho_{xx}(t,v)
 \end{pmatrix}
 \end{equation}
 and
 \begin{equation}
 \bar\rho (t,v) = \begin{pmatrix}
 \bar\rho_{ee}(t,v) & \bar\rho_{ex}(t,v)\\ 
 \bar\rho_{ex}^*(t,v)&\bar\rho_{xx}(t,v)
 \end{pmatrix}\ ,
\end{equation}
 we can linearize the equations of motion at first order in $|\rho_{ex}|$, if $|\rho_{ex}|\ll|\rho_{ee}-\rho_{xx}|$. 
 
 The linearized equations of motion for the off-diagonal element of the density matrix are
 \begin{widetext}
\begin{eqnarray}
\label{eq:Linear_Equations}
 i\partial_t \rho_{ex}(t,v) &=& (H_{ee}-H_{xx})\rho_{ex}(t,v) - g_v \left[\frac{\omega^s}{2}+ \mu\int \mathrm{d}v'(1-vv')(\rho_{ex}(t,v^{\prime})-\bar\rho_{ex}(t,v^{\prime})) \right]\ , \nonumber\\ 
i\partial_t \bar{\rho}_{ex}(t,v) &=& (\bar{H}_{ee}-\bar{H}_{xx})\bar{\rho}_{ex}(t,v) + \bar{g}_v \left[\frac{\omega^s}{2}- \mu\int \mathrm{d}v'(1-vv')(\rho_{ex}(t,v^{\prime})-\bar\rho_{ex}(t,v^{\prime})) \right]\ ,\nonumber\\
\end{eqnarray}
\end{widetext}
where $g_v = \rho^0_{ee}(v)-\rho^0_{xx}(v)$ and $\bar g_v = \bar\rho^0_{ee}(v)-\bar\rho^0_{xx}(v)$, respectively  and with $0$ denoting $t=0$. 
The term proportional to $\omega^s$ is the flavor-violating component of the vacuum Hamiltonian responsible for providing the initial seed when $\rho_{ex} = 0$. However, once the initial seed grows, the term proportional to $\mu$ takes over. Since $\mu\gg\omega$, we can neglect the term proportional to $\omega^s$. Assuming a collective regime in which all neutrino and antineutrino modes evolve with the same eigenfrequency $\Omega=\gamma+i\kappa$, we can express the off-diagonal terms of the density matrix through the following plane-wave ansatz:
\begin{equation}
\label{eq:LI_Solutions}
 \rho_{ex}(t,v) = Q_v e^{-i\Omega t} \;\;\;\;\; \mathrm{and} \;\;\;\;\; \bar\rho_{ex}(t,v) = \bar Q_v e^{- i\Omega t}\ .
\end{equation}
When the eigenfrequency has a non-zero imaginary part (i.e., $\kappa\neq0$), a flavor instability exists. Substituting the ansatz above in the linearized equations of motion and assuming that $\accentset{(-)}{H}_{ee}-\accentset{(-)}{H}_{xx} \simeq \rm{const.}$, we have
\begin{subequations}
 \begin{equation}
 [\Omega - (H_{ee}-H_{xx})] Q_{v} = -g_v \mu \int \mathrm{d} v' (Q_{v'}-\bar Q_{v'}) (1-v v')\ ,
 \end{equation}
 \begin{equation}
 [\Omega - (\bar H_{ee}-\bar H_{xx})] \bar Q_{v} = -\bar{g}_v \mu \int \mathrm{d} v' (Q_{v'}-\bar Q_{v'} ) (1-v v')\ .
 \end{equation}
\end{subequations}
The equations above lead to a solution that is  polynomial in $v$:
\begin{equation}
\label{eq:Q_v_ansatz}
 Q_v = -g_v\frac{(a-bv)}{\Omega- (H_{ee}-H_{xx})}
 \end{equation}
 \begin{equation}
 \bar Q_v = -\bar{g}_v \frac{(a-bv)}{\Omega- (\bar H_{ee}-\bar H_{xx})}\ . \nonumber
\end{equation}
Plugging the expressions above in the equations of motion, we obtain a system of equations, for which non-trivial solutions are found if
 \begin{equation}
 \label{eq:Det_0}
 \det[D(\Omega)]=
 \begin{vmatrix}
 I_0-1 & -I_1 \\ I_1 & -I_2 -1
 \end{vmatrix}= 0\ ,
 \end{equation}
 with
 \begin{equation}
 \label{eq:In}
 I_n(\Omega) = \mu \int \mathrm{d} v v^n \left [ \frac{\bar g_v}{\Omega- (\bar H_{ee}-\bar H_{xx})} -\frac{g_v}{\Omega- (H_{ee}-H_{xx})}\right]\ .
 \end{equation}We can further expand the denominator of $I_n(\Omega)$, using 
\begin{subequations}
 \begin{equation}
  H_{ee}-H_{xx} = - \omega^c +\lambda + \mu (D^z_0 - v D^z_1)\ ,  
\end{equation}
\begin{equation}
  \bar{H}_{ee}-\bar{H}_{xx} = +\omega^c +\lambda + \mu (D^z_0 - v D^z_1)\ ,  
\end{equation}
\end{subequations}
where $D^z_0$ and $D^z_1$ are the initial $z$-components of the zeroth and first angular moments of $\vec D_n$. It is clear that the Hamiltonians of neutrinos and antineutrinos share the same dependence on $\lambda$ and $D^z_0$ (the latter is constant given that $D_0^z=D_0^z|_{t=0} + \mathcal{O}(\rho_{ex}^2)$). These terms only affect the real part of $\Omega$, as shown in Appendix~\ref{sec:matter}.

Since we are interested in looking for instabilities, these diagonal potentials can be absorbed in $\Omega'=\Omega-\lambda-\mu D^z_0$, so that 
\begin{equation}
\label{eq:I_f_theta}
 I_n(\Omega') = \mu \int \mathrm{d}v v^n \left [ \frac{\bar g_v}{\Omega'+ \mu vD^z_1 -\omega^c} -\frac{g_v}{\Omega'+\mu vD^z_1 +\omega^c}\right]\ .
\end{equation}
This equation is important to explore the role of $\omega\neq0$ in the development of flavor instabilities that otherwise develop for $\omega \rightarrow 0$, as discussed in the following sections. Note that $\omega^c$ has a different sign in the terms proportional to $g_v$ and $\bar{g}_v$, which is key in defining the role of $\omega$ in the stability of this system. 

\subsection{Parametrization of the neutrino angular distributions}
\label{sec:Initial_Angular_Dist}
We model the angular distributions of neutrinos and antineutrinos through Gaussians distributions centered on $v=1$ and standard deviation $\sigma_{\nu_\beta}$:
\label{eq:rho_Gaussian_Param}
 \begin{equation}
 \rho^0_{ee} (v)=\mathcal{G}(v;1,\sigma_{\nu_e})
 \end{equation}
 normalized such that
 $\int_{-1}^{+1} \mathrm{d}v \rho^0_{ee} (v) = 1$,
 \begin{equation}
 \bar \rho^0_{ee}(v) = \alpha_{\bar\nu_e} \mathcal{G}(v;1,\sigma_{\bar\nu_e})
  \end{equation}
 with 
$ \int_{-1}^{+1} \mathrm{d}v \bar\rho^0_{ee} (v) = {n_{\bar\nu_e}}/{n_{\nu_e}} =\alpha_{\bar\nu_e}$,
 \begin{equation}
 \rho^0_{xx}(v)= \alpha_{\nu_x} \mathcal{G}(v;1,\sigma_{\nu_x})
  \end{equation}
where
$\int_{-1}^{+1} \mathrm{d}v \rho^0_{xx} (v)= {n_{\nu_x}}/{n_{\nu_e}} = \alpha_{\nu_x}$.
In our distributions,  $\alpha_{\nu_{\beta}}$ is a scaling factor directly linked to the relative abundance of neutrinos of flavor $\beta$ with respect to $\nu_e$. We consider an ensemble of angular distributions with $\alpha_{\bar \nu_e}\in\left[0.4,1.2\right]$ and $\sigma_{\bar \nu_e}\in\left[0.1,1.0\right]$ for $\bar\nu_e$, while the other parameters are  fixed as indicated in Table~\ref{tab:Benchmark}.

\begin{figure*}[tbp]
\centering 
\includegraphics[width=0.49\textwidth]{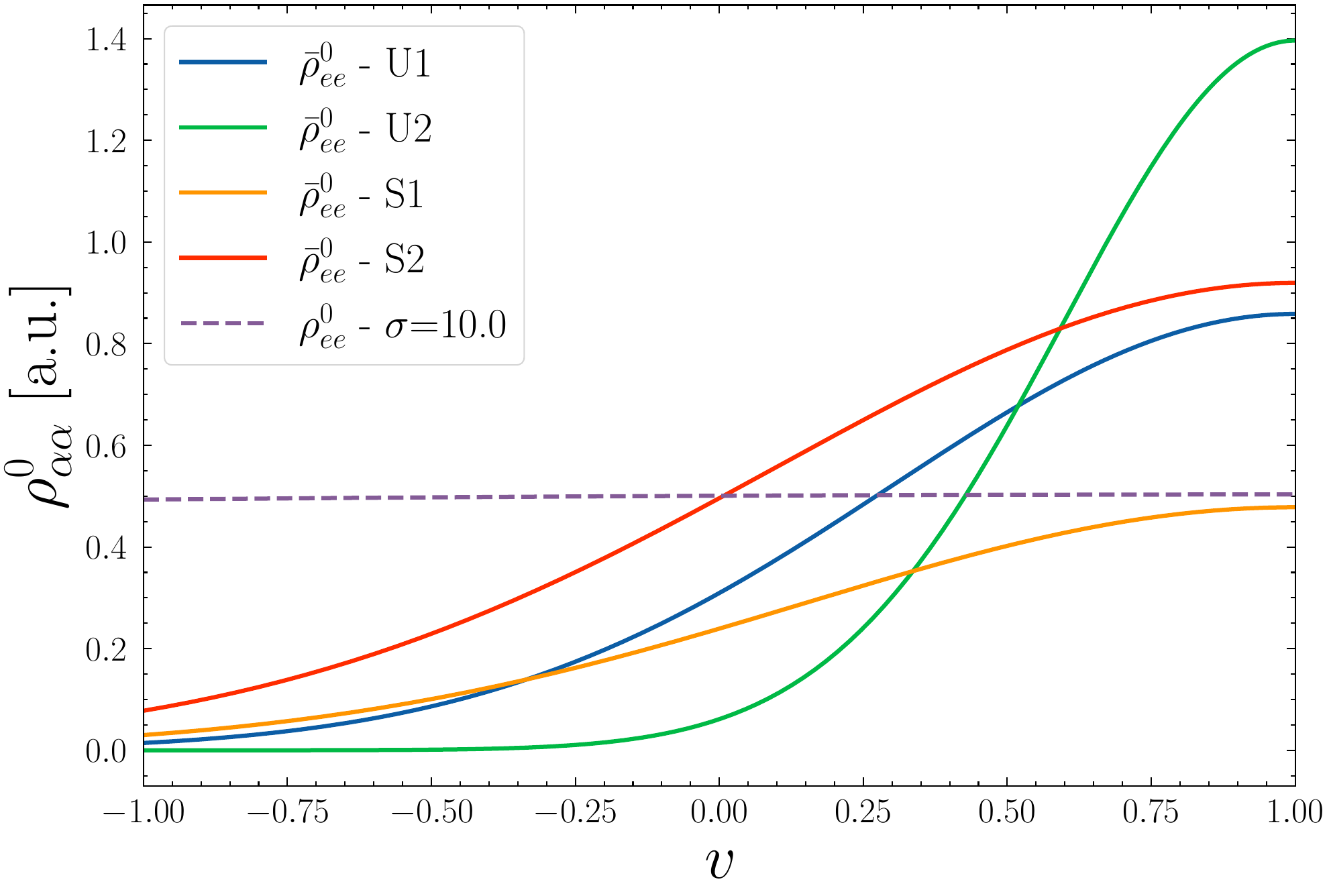}
\includegraphics[width=0.49\textwidth]{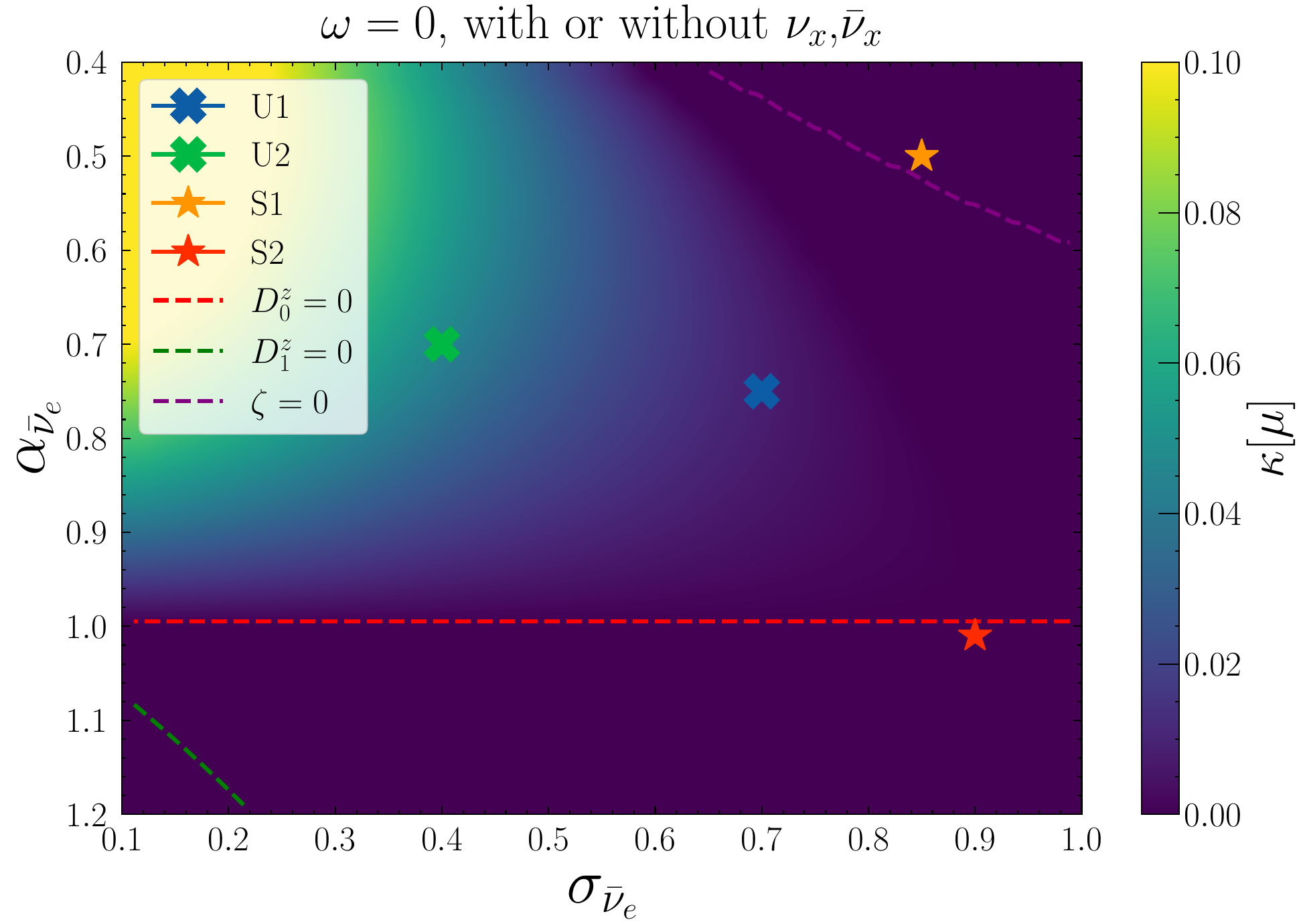}
\caption{{\it Left:} Initial angular distributions for the four benchmark cases adopted in this work (see also Table~\ref{tab:Benchmark}). {\it Right:} Contour plot of the normalized growth rate for vanishing vacuum mixing ($\omega=0$) in the plane spanned by $\sigma_{\bar\nu_{e}}$ and $\alpha_{\bar\nu_{e}}$, for $\sigma_{\nu_e}=10$. Our four benchmark cases are highlighted, together with the isocontours for the ELN number $D^z_0=0$ and flux $D^z_1=0$, and $\zeta=0$ (see text for details). If $D^z_0$ and $D^z_1$ have opposite signs, the pendulum is stable. If $\zeta>0$, an ELN zero-crossing exists, which is a necessary condition for the existence of fast flavor instabilities.}
\label{fig:Alpha_times_Sigma_and_Angular_Dist}
\end{figure*}

The left panel of Fig.~\ref{fig:Alpha_times_Sigma_and_Angular_Dist} shows four representative angular distributions that we use as benchmark cases in our work and are comparable to the ones explored in Ref.~\cite{Padilla-Gay:2021haz}: cases U1, U2, S1, and S2, whose characteristic parameters are summarized in Table~\ref{tab:Benchmark}. As we show in Sec.~\ref{sec:Vanishing_w}, cases U1 and U2 are unstable in the fast limit ($\omega\rightarrow 0$), while cases S1 and S2 are stable for $\omega\rightarrow 0$. The goal of this work is to explore how the instability regions that were investigated in Ref.~\cite{Padilla-Gay:2021haz} for $\omega=0$ are modified by $\omega \neq 0$.

\begin{table}
\centering
\caption{Input values for the four selected angular distributions of $\nu_e$, $\bar\nu_e$ and $\nu_x$ considered as benchmark cases in this work. \label{tab:Benchmark}}
\vspace{.5cm}
\begin{tabular}{|c|c|c|c|c|c|}
 \hline
Case & $\sigma_{\nu_e}$ & $\alpha_{\bar\nu_e}$ & $\sigma_{\bar\nu_e}$ & $\alpha_{\nu_x}$ & $\sigma_{\nu_x}$/$\sigma_{\bar\nu_e}$ \\ 
 \hline
 \hline
 U1 & 10 & 0.75 & 0.70 &1.0 &0.4\\
 U2 & 10 & 0.70 & 0.40 &1.0 &0.4\\
 S1 & 10 & 1.01 & 0.90 &1.0 &0.4\\
 S2 & 10 & 0.50 & 0.85 &1.0 &0.4 \\
 \hline
\end{tabular}
\end{table}

\section{Flavor instability for vanishing vacuum mixing}
\label{sec:Vanishing_w}
Before exploring the impact of $\omega$ in the development of flavor instabilities, we review the stability conditions for our system in the limit of $\omega \rightarrow 0$. Equation~\ref{eq:I_f_theta} can be rewritten as 
\begin{equation}
\label{eq:I_f_theta_w_0}
 I_n(\Omega') = \mu \int \mathrm{d}v v^n \frac{(\bar g_v- g_v)}{\Omega'+ \mu vD^z_1}= \mu \int \mathrm{d}v v^n \frac{(\bar \rho^0_{ee}- \rho^0_{ee})}{\Omega'+ \mu vD^z_1}\ ,
\end{equation}
where we assume $\rho^0_{xx}=\bar{\rho}^0_{xx}$. The equation above highlights the dependence on the ELN angular distribution, given by $\rho^0_{ee}- \bar{\rho}^0_{ee}$~\cite{Chakraborty:2016lct,Izaguirre:2016gsx,Morinaga:2021vmc}. 

The right panel of Fig.~\ref{fig:Alpha_times_Sigma_and_Angular_Dist} displays the growth rate of the flavor instability ($\kappa$) in the plane spanned by $\alpha_{\bar\nu_e}$ and $\sigma_{\bar\nu_e}$. For completeness, in the same plot, we also show three quantities of relevance to interpret the stability conditions of our system: the loci of vanishing lepton number ($D^z_0 =0$) and flux ($D^z_1=0$)~\cite{Padilla-Gay:2021haz,Johns:2019izj,Fiorillo:2023hlk,Fiorillo:2023mze}, and the ELN zero-crossing parameter $\zeta$. The latter is defined as~\cite{Shalgar:2019qwg}
 \begin{equation}
 \zeta=\frac{I_1I_2}{(I_1+I_2)^2}\ ,
  \end{equation} 
  with
   \begin{equation}
 I_1= \int \mathrm{d} v\ \Theta(\rho^0_{ee}-\bar\rho^0_{ee})\;\;\;\; \mathrm{and} \;\;\;\;I_2=\int \mathrm{d} v\ \Theta(\bar\rho^0_{ee}-\rho^0_{ee})\ ;
 \end{equation} 
 $\Theta$ is the Heavyside function, resulting in $\zeta>0$, if an ELN zero-crossing exists. The existence of an ELN zero-crossing is necessary for the development of a fast flavor instability~\cite{Padilla-Gay:2021haz,Morinaga:2021vmc,Dasgupta:2021gfs}. Hence the region above $\zeta=0$ in the right panel of Fig.~\ref{fig:Alpha_times_Sigma_and_Angular_Dist} is stable. 
 
 This system is formally equivalent to a gyroscopic pendulum~\cite{Padilla-Gay:2021haz,Johns:2019izj}, with $\vec{D}_0$ and $|\vec D_1|$ being conserved. Within this analogy, $\vec D_0$ plays the role of gravity, and $\vec D_1$ represents the center-of-mass position with respect to the point of support. If $\vec{D}_0$ and $\vec{D}_1$  have opposite signs, the pendulum is initially aligned with the gravitational field and it is therefore stable. The right panel of Fig.~\ref{fig:Alpha_times_Sigma_and_Angular_Dist}  shows that our benchmark cases S1 and S2 sit in a region of the parameter space where the neutrino system is stable. S1 is stable because it does not exhibit an ELN zero-crossing, while S2 belongs to a region of the parameter space where the pendulum starts in a stable position. On the other hand, cases U1 and U2 are unstable in the limit of $\omega \rightarrow 0$. 

\section{Flavor instability for non-vanishing vacuum mixing}
\label{sec:Non_Vanishing_w}

In this section, we intend to investigate how the flavor unstable region in the right panel of Fig.~\ref{fig:Alpha_times_Sigma_and_Angular_Dist} is modified by the presence of $\omega \neq 0$. 
To this purpose, first we consider a few examples of ELN  angular distributions that are unstable despite the absence of ELN zero-crossings in the angular distribution.
Then, we outline the role of $\omega$ through a perturbative expansion and explore the role of the linear corrections due to $\omega$ for our four benchmark cases. We also illustrate how the effect of $\omega$ can be taken into account by introducing effective angular distributions in an equivalent system with $\omega=0$; this approach is especially useful to explain why ELN configurations that are stable for $\omega \rightarrow 0$ become unstable for $\omega \neq 0$.

\subsection{Examples of unstable ELN configurations without zero-crossing} \label{sec:example}
\begin{figure*}
\includegraphics[width=0.49\textwidth]{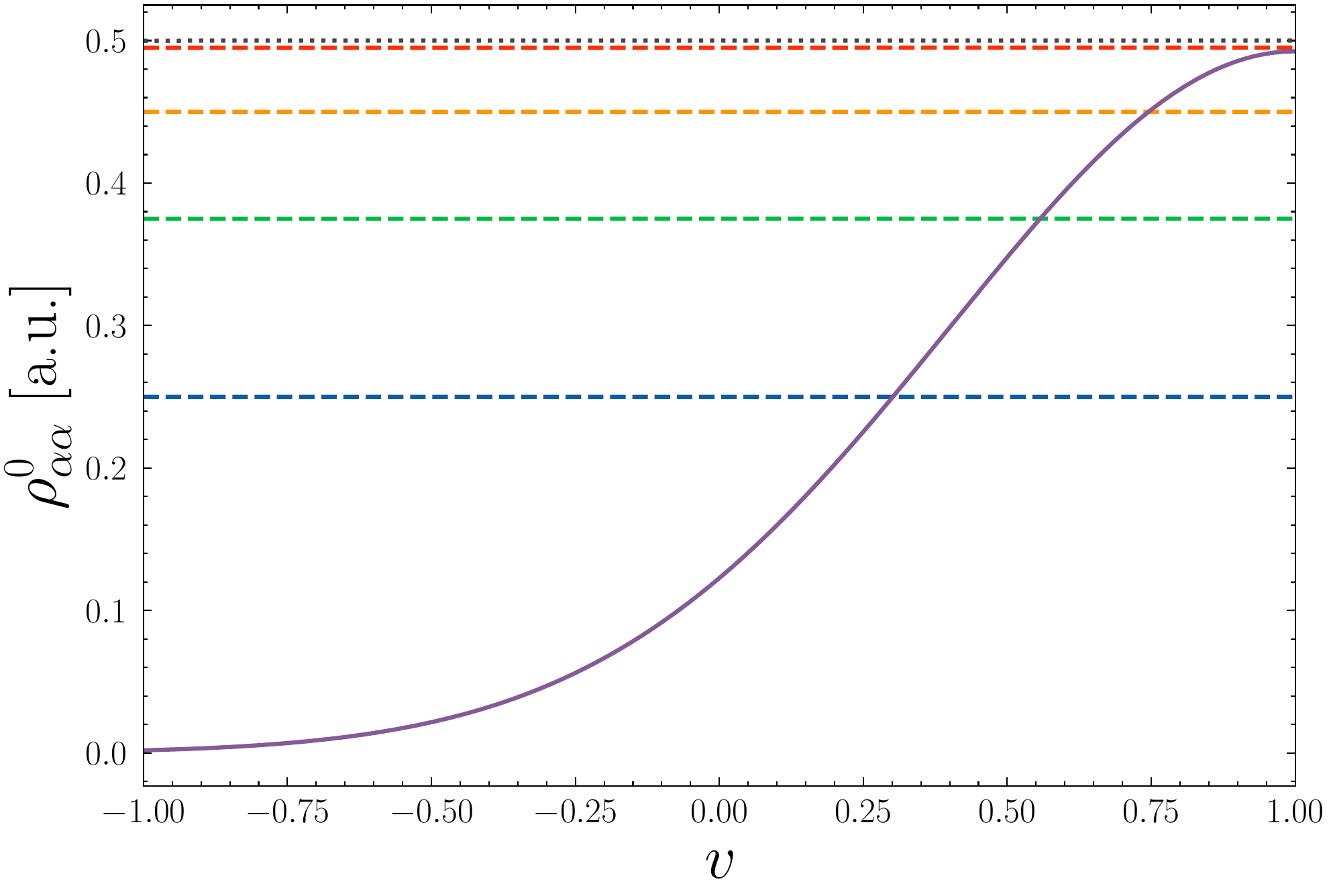}
\includegraphics[width=0.49\textwidth]{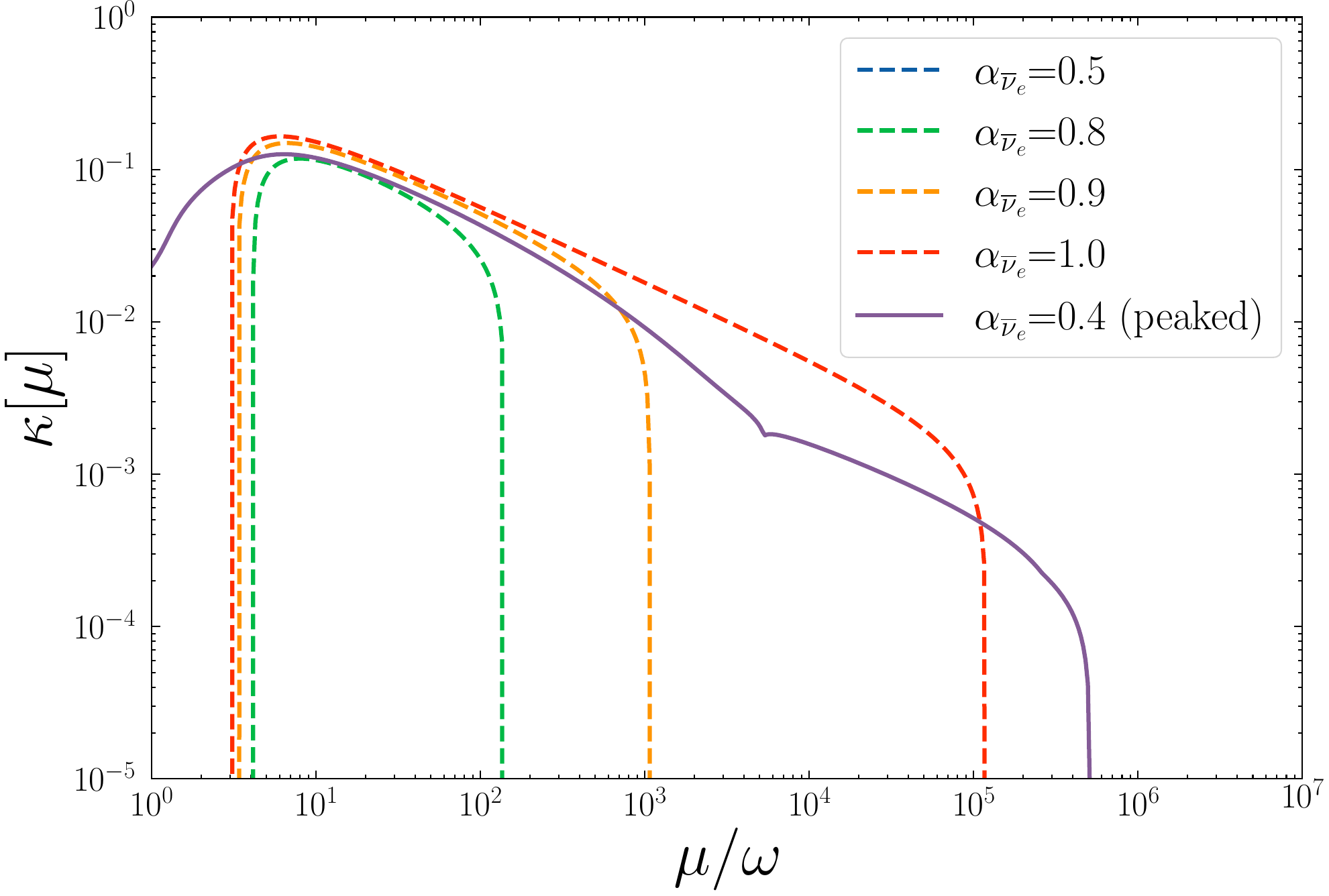}
		\caption{{\it Left:} Angular distribution of $\nu_{e}$ (black dotted line) and $\bar{\nu}_{e}$ (dashed lines with different colors). The dashed lines show  $\bar{\nu}_{e}$ isotropic configurations with different normalizations;  the forward peaked angular distribution of $\bar{\nu}_{e}$ (solid indigo line) has an angle integrated normalization of $0.4$. {\it Right:} Growth rate of the off-diagonal term of the density matrix obtained using the linear stability analysis for the ELN configurations shown in the left panel. For the configurations with isotropic distributions for both electron flavors, as $\alpha_{\bar{\nu}_{e}}$ increases the instability spans a larger range of $\mu/\omega$. The flavor configuration with the forward peaked $\bar\nu_e$ distribution has the largest range of $\mu/\omega$ over which the flavor instability exists.}
		\label{growthrates}
\end{figure*}

We first consider a few simpler examples with isotropic angular distributions for  $\bar\nu_e$, as illustrated in the left panel of Fig.~\ref{growthrates} (cf.~dashed lines). While keeping unchanged the angular distribution of $\nu_e$, we also consider a forward peaked angular distribution for $\bar\nu_e$ (cf.~solid line in the left panel of Fig.~\ref{growthrates}) with $\sigma_{\bar{\nu}_e}=0.6$ and $\alpha_{\bar{\nu}_e}=0.4$. In none of the considered configurations an ELN zero-crossing is present. 

The right panel of  Fig.~\ref{growthrates} displays the growth rate of the flavor instability for the configurations considered in the left panel of the same figure as a function of $\mu/\omega$. 
It is clear that the range of $\mu/\omega$ over which the flavor instability exists is the largest for the system with a forward peaked distribution. Moreover, for the configurations with isotropic distributions of both $\nu_e$ and $\bar\nu_e$, as $\alpha_{\bar{\nu}_{e}}$ increases the  instability  spans a larger range of $\mu/\omega$. Hence, the flavor instability does not only depend on $\alpha_{\bar{\nu}_{e}}$, $\mu$ and $\omega$, but also on the shape of the angular distribution.

Relying on the neutrino-bulb model, it was concluded the system is stable in regions of high neutrino density, unless an ELN zero-crossing is present~\cite{Esteban-Pretel:2008ovd,Esteban-Pretel:2007jwl,Chakraborty:2011nf}. However, Fig.~\ref{growthrates} shows that, despite the large neutrino density and the absence of an ELN zero-crossing, a flavor instability can exist. In the following section,  we investigate the neutrino equations of motion in the linear regime with a perturbative expansion in the vacuum term to gain insight into the reason for this behavior.

\subsection{Perturbative expansion}
The vacuum term of the Hamiltonian has opposite signs for neutrinos and antineutrinos. Because of this, Eq.~\ref{eq:I_f_theta} does not have a simple expression in terms of $\bar{g}_v - g_v$. However, one can gain insight into the modifications introduced by $\omega$ through a perturbative expansion around $\omega=0$. To illustrate this, let us consider the neutrino part of the integral of Eq.~\ref{eq:I_f_theta}:
\begin{equation}
\begin{split}
 I^\nu_n(\Omega') &= \mu \int \mathrm{d}v v^n \frac{g_v}{\Omega'+ \mu v D^z_1} \left(1+\frac{\omega^c}{\Omega'+\mu vD^z_1}\right)^{-1}\\ &= \mu \int \mathrm{d}v v^n \frac{g_v}{\Omega'+ \mu v D^z_1} \sum_{k=0}^{\infty} \left(\frac{-\omega^c}{\Omega'+\mu vD^z_1}\right)^{k}\ .
\end{split}
\end{equation}
A similar expansion can be carried out for the antineutrino part of Eq.~\ref{eq:I_f_theta} replacing $\omega^c \rightarrow -\omega^c$. Combining the neutrino and antineutrino contributions, Eq.~\ref{eq:I_f_theta} becomes 
\begin{equation}
\label{eq:I_f_theta_expansion}
 I_n(\Omega') = \mu \sum_{k=0}^{\infty} \int \mathrm{d}v v^n \frac{\bar{g}_v-(-1)^k g_v}{\Omega'+\mu v D^z_1} \left(\frac{\omega^c}{\Omega'+\mu vD^z_1}\right)^{k}\ .
\end{equation}
The structure of Eq.~\ref{eq:I_f_theta_expansion} is similar to the one of Eq.~\ref{eq:I_f_theta_w_0}, which is fully recovered when considering the zeroth-order term only. However, for higher-order corrections, the odd $k$ terms depend on $(\bar{g}_v+ g_v)$, rather than on $(\bar{g}_v-g_v)$. If we separate the spectra in their flavor components, we find that the odd $k$ corrections depend on the neutrino flavor particle number (FPN, given by $\rho^0_{\alpha\alpha}+\bar \rho^0_{\alpha\alpha}$), instead of the neutrino flavor lepton number (FLN,  given by~$\rho^0_{\alpha\alpha}-\bar \rho^0_{\alpha\alpha}$). As a consequence, a dependence on the angular distributions of $\nu_x$ and $\bar\nu_x$ appears, which is not present in the limit of $\omega \rightarrow 0$ (see also the right panel of Fig.~\ref{fig:Alpha_times_Sigma_and_Angular_Dist}). A similar procedure can be followed for the growth rate $\kappa$, developing a Taylor expansion around $\omega=0$:
\begin{equation}
\label{eq:kappa_Taylor_w_0}
 \kappa(\omega) = \sum_{k=0}^{\infty} a_k \omega^k= a_0+a_1 \omega + a_2 \omega^2 +\mathcal{O}(\omega^3)\ ,
\end{equation}
where the coefficients $a_k$ are to be determined, relying on the behavior of $I_{n}(\Omega')$. 

\subsection{First order correction}
\label{sec:Linear_in_w}
 Since $\omega \ll \mu$,  we expand Eq.~\ref{eq:I_f_theta_expansion} up to the first order in $\omega$:
\begin{widetext}
\begin{eqnarray}
\label{eq:I_f_theta_first_order}
 I_n(\Omega') = \mu \int \mathrm{d} v v^n \left[\frac{(\bar\rho^0_{ee}-\rho^0_{ee})-(\bar\rho^0_{xx}-\rho^0_{xx})}{\Omega'+\mu v D^z_1} +
 +\omega^c \frac{(\bar\rho^0_{ee}+\rho^0_{ee})-(\bar\rho_{xx}+\rho_{xx})}{(\Omega'+\mu v D^z_1)^2}\right]\ . 
\end{eqnarray}
\end{widetext}
The dependence of the equation above on the difference EPN-XPN is canceled out if $\bar\rho^0_{xx}+\rho^0_{xx} = \bar\rho^0_{ee}+\rho^0_{ee}$. In this case, there is no first-order correction to the stability equations (i.e., $a_1=0$ in Eq.~\ref{eq:kappa_Taylor_w_0}). 

\begin{figure*}
\includegraphics[width=0.49\textwidth]{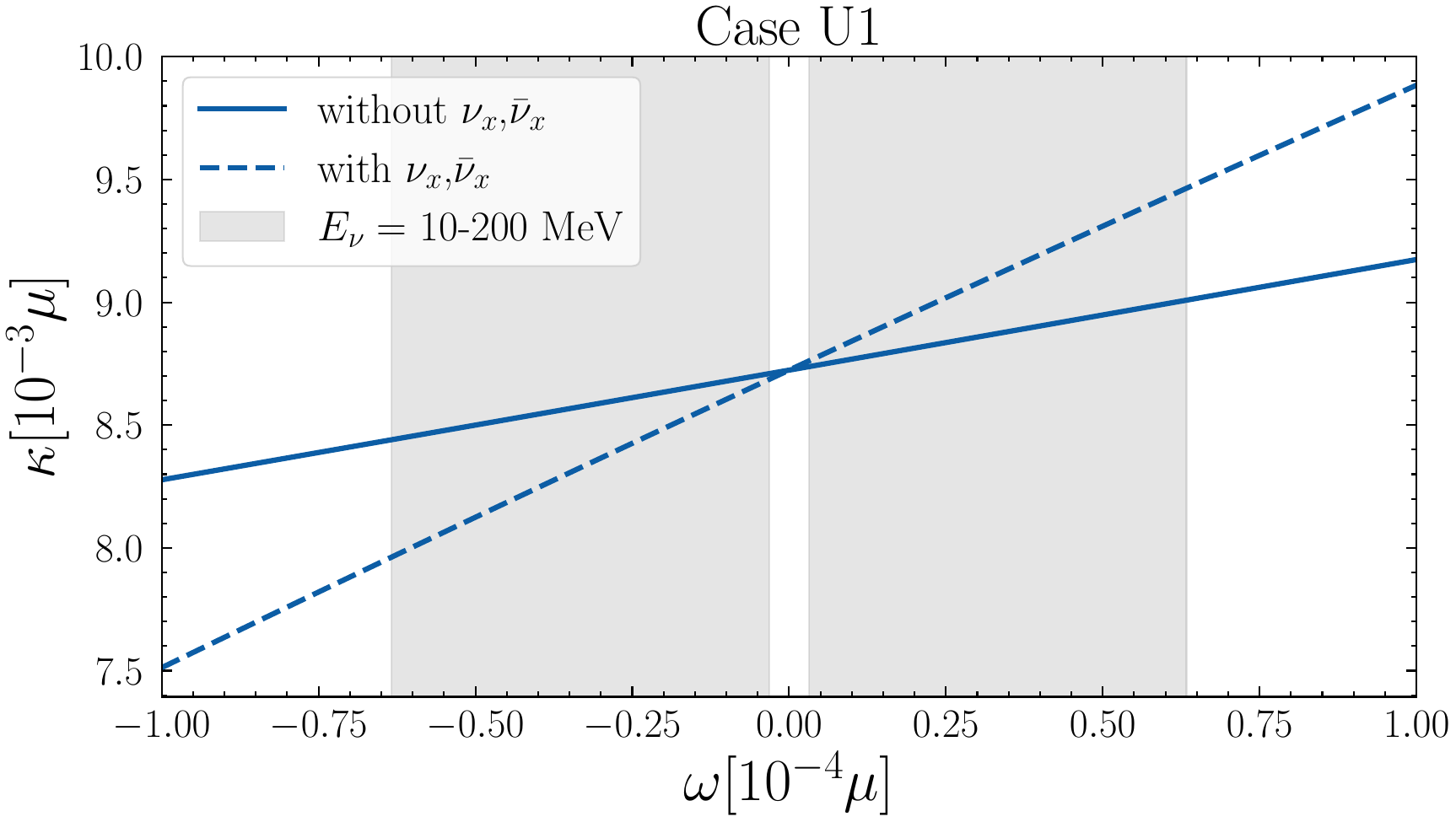}
\includegraphics[width=0.49\textwidth]{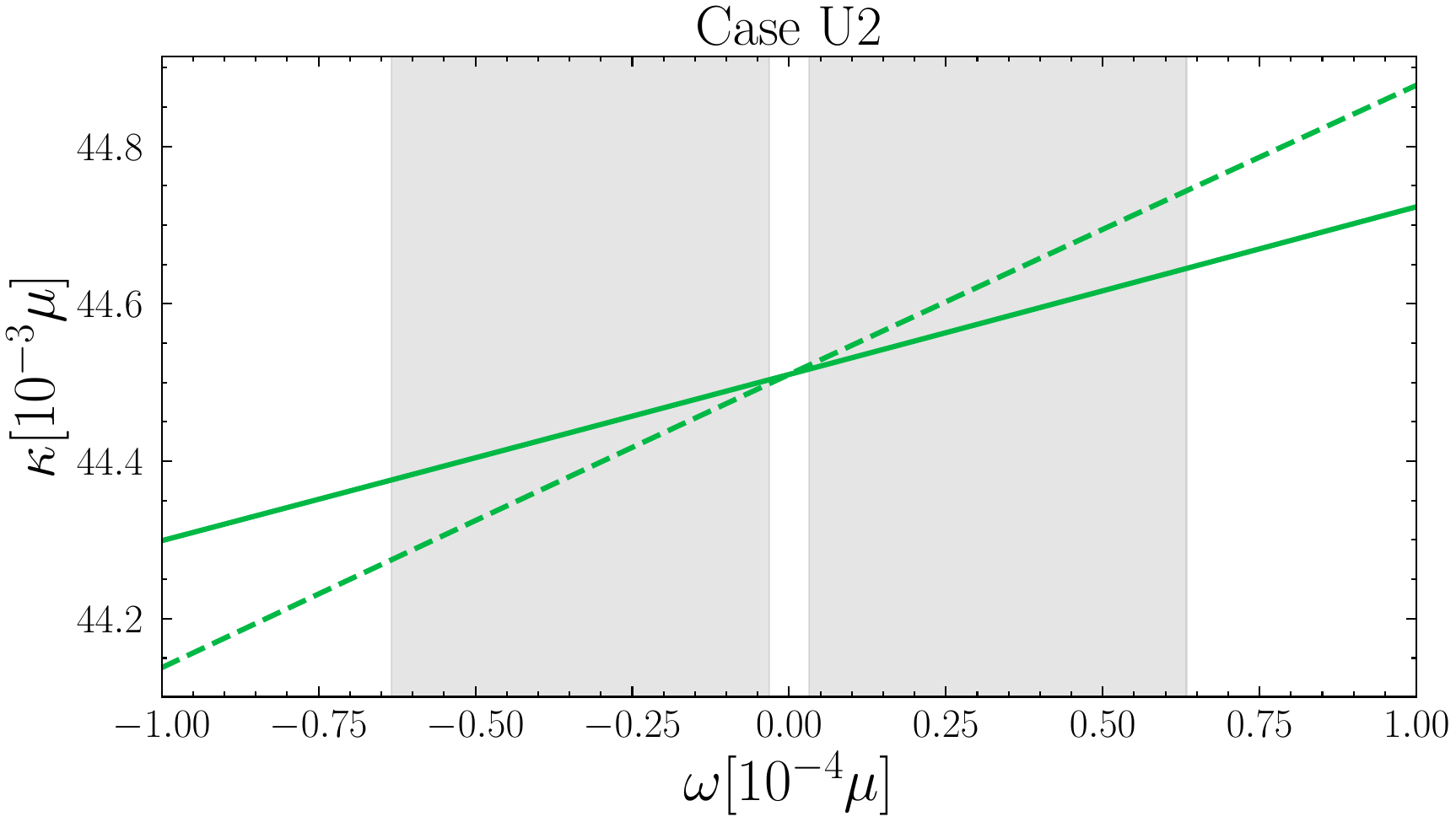}
\includegraphics[width=0.49\textwidth]{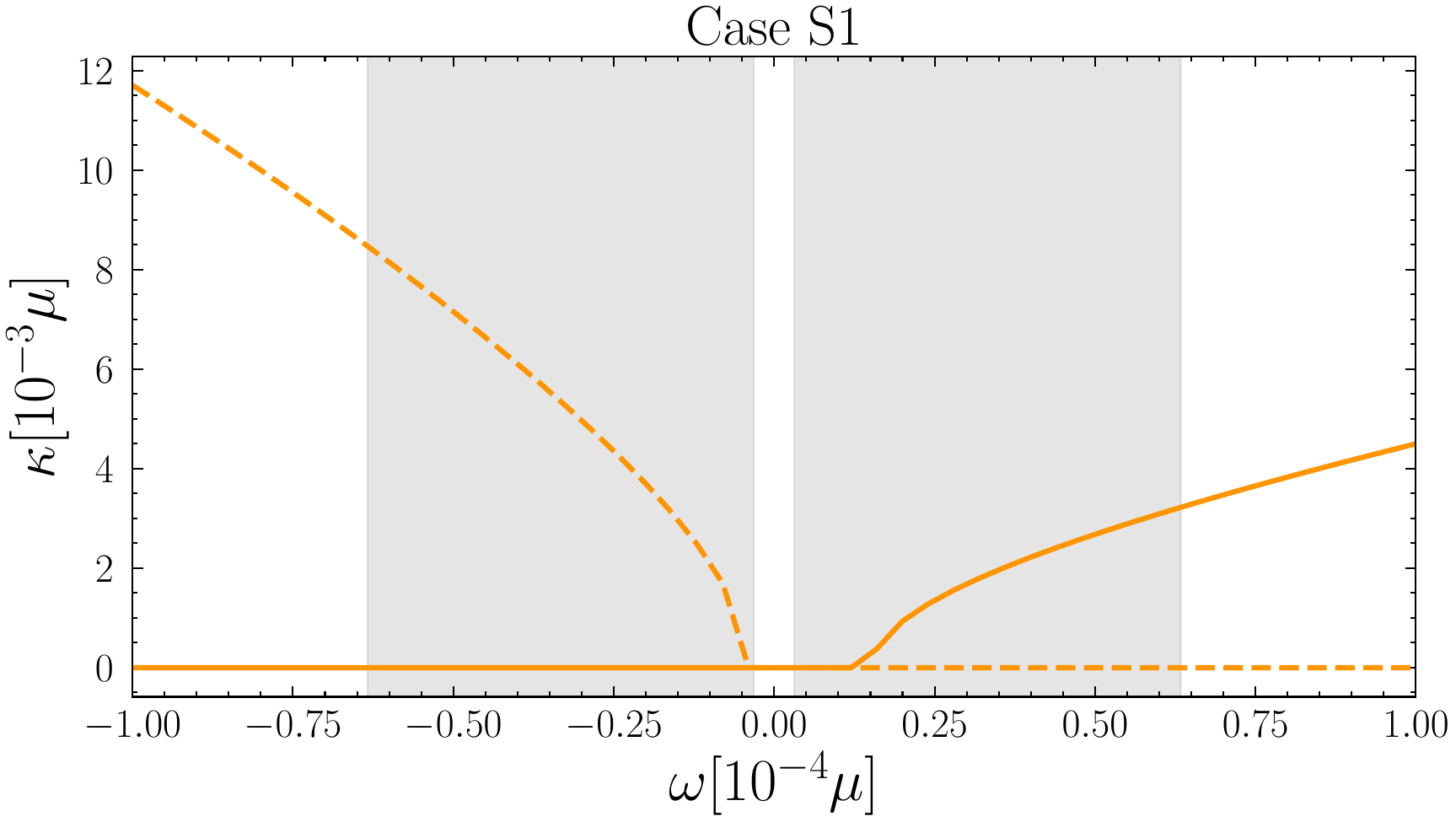}
\includegraphics[width=0.49\textwidth]{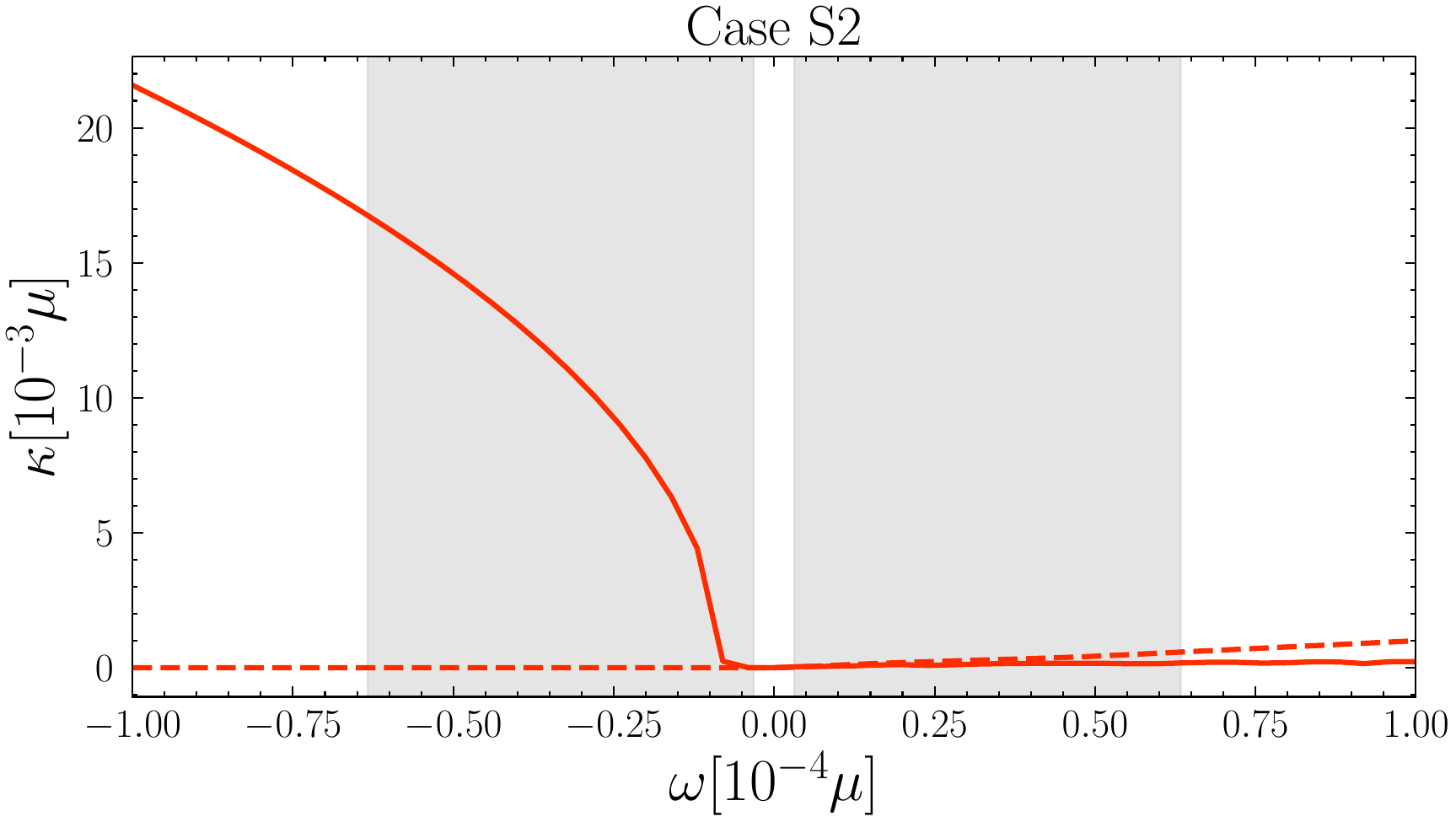}
\caption{Growth rate $\kappa$ as a function of $\omega$ with (dashed) and without (solid) non-electron neutrinos. The top (bottom) panels correspond to configurations U1 and U2 (S1 and S2). The gray shaded areas correspond to vacuum frequencies for $E=10$--$100$~MeV, considering $\mu=10^{4}$ km$^{-1}$ and $\Delta m^2=\Delta m^2_{31} = 2.5 \times 10^{-3}$ eV$^{2}$, and $\theta_V=10^{-6}$. Both U1 and U2 show a linear growth of $\kappa$ with $\omega$, while S1 and S2 develop instabilities for $\omega\neq0$ with an asymmetric behavior around $\omega=0$, which depends on the angular distribution of non-electron neutrinos. }
\label{fig:Kappa_x_omega}
\end{figure*}

In a neutrino gas with $\omega \neq 0$,   a linear dependence on $\omega$ appears, as one can see in the top panels of Fig.~\ref{fig:Kappa_x_omega} for cases U1 and U2~\footnote{Although $a_1$ in Eq.~\ref{eq:kappa_Taylor_w_0} has a non-trivial dependence on the angular distribution, one can see that the inclusion of $\nu_x$ leads to a larger $a_1$. We have checked that, more generally, $a_1$ is positive if $\sigma_{\nu_x}$ is below a critical value and negative otherwise, where the critical value depends on the angular distribution of the EPN. However, for  $\sigma_{\nu_e}>\sigma_{\bar\nu_e}>\sigma_{\nu_x}$, as expected in the case of core-collapse supernovae, $a_1$ is positive.}. 
The growth rate obtained for $\omega \neq 0$ is of the same order as the one for $\omega=0$. Importantly, for the neutrino configurations that are stable in the limit $\omega \rightarrow 0$, flavor instabilities develop for $\omega \neq 0$, as discussed in the next section.

Note that the dependence on the FPN was pointed out in the context of slow flavor conversion. Reference~\cite{Raffelt:2007yz} explicitly demonstrated that the flavor evolution depends on the coupling between the lepton number density ($D_{v} = \rho_{v} - \bar\rho_{v}$) and the particle density ($S_{v} = \rho_{v} + \bar\rho_{v}$) because of the vacuum term in the Hamiltonian. On the other hand, the fast flavor conversion dynamics is driven by the energy-integrated lepton number ($D_{\vec{p}}$)~\cite{Padilla-Gay:2021haz}. The fact that we consider $\omega \neq 0$ couples the evolution of $D_{\vec{p}}$ and $S_{\vec{p}}$, similarly to what was found in Ref.~\cite{Raffelt:2007yz}. 

\subsection{Neutrino flavor instabilities without non-electron flavors}
\label{sec:Slow_Instabilities}
We now explore the role of $\omega$ in the development of flavor instabilities, first focusing on the case without non-electron flavors. The bottom panels of Fig.~\ref{fig:Kappa_x_omega} display the instability growth rate as a function of $\omega$ for the neutrino configurations S1 and S2. 
One can see that $\kappa \simeq 0$ for $\omega=0$ (we use $\kappa_{\text{min}}=10^{-10}\mu$ as a threshold to mark the presence of flavor instabilities, see Appendix~\ref{sec:Appendix_Numerical}), but this is not the case for $\omega \neq 0$ (similar to the preliminary findings reported in Sec.~\ref{sec:example}).

Equation~\ref{eq:I_f_theta_first_order} also suggests that the sign of $\omega$ has different implications on the growth rate: it can either contribute to the development of instabilities or have the opposite effect, according to the shape of the angular distributions. 
In fact, Fig.~\ref{fig:Kappa_x_omega}  shows an asymmetric behavior around $\omega=0$. For the configurations without $\nu_x$ (solid curves), the flavor instability develops for $\omega>0$ ($\omega<0$) for case S1 (S2). This asymmetry is directly related to the different origin of the flavor stability: case S1 does not have an ELN zero-crossing (but it effectively develops one because of the vacuum term), while case S2 presents an ELN zero-crossing, but it has opposite signs for $D^z_0$ and $D^z_1$. 

\begin{figure*}
\includegraphics[width=0.48\textwidth]{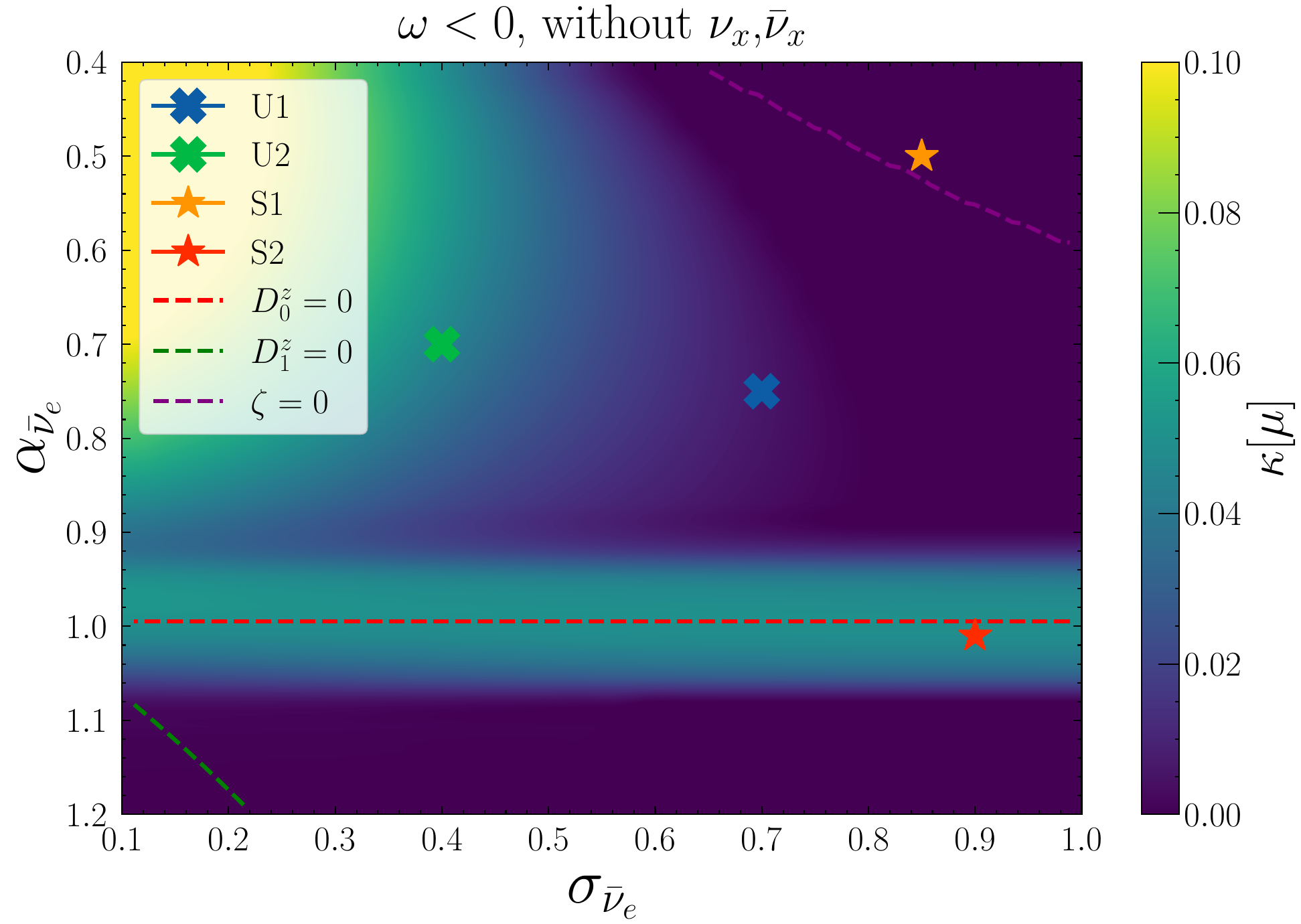}
\includegraphics[width=0.48\textwidth]{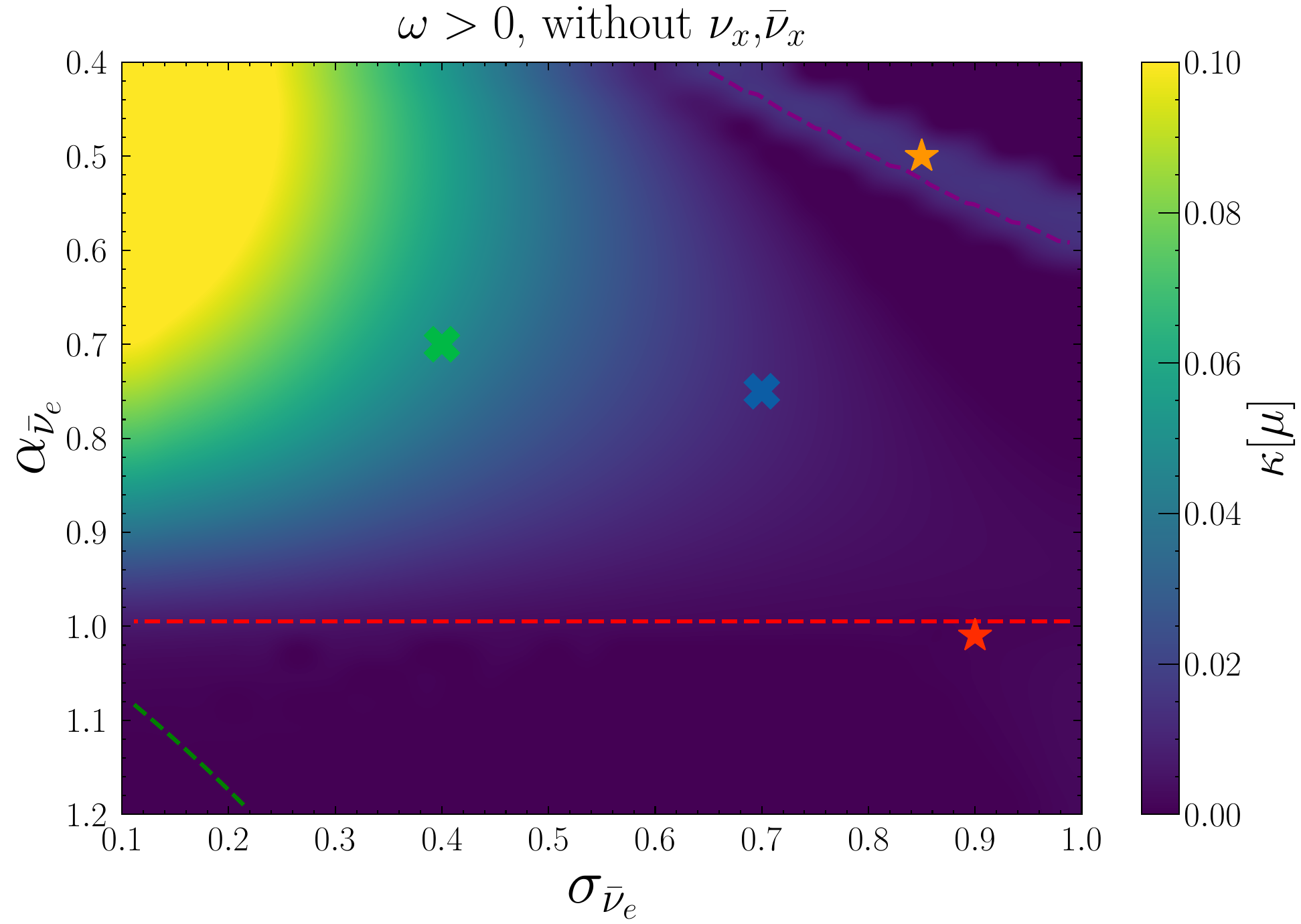}
\caption{Contour plot of the growth rate $\kappa$ in the plane spanned by $\sigma_{\bar\nu_{e}}$ and $\alpha_{\bar\nu_{e}}$ for the scenario without $\nu_x,\bar\nu_x$ and $\omega=-5\times10^{-4}\mu$ (left) and $\omega=5\times10^{-4}\mu$ (right). To guide the eye, our benchmark configurations U1, U2, S1, and S2 are marked with a cross in each plot, while the isocontours of $D^z_0=0$, $D^z_1=0$, and $\zeta=0$ (see text for details) are plotted as dashed lines. As it can be noted comparing this figure with the right panel of Fig.~\ref{fig:Alpha_times_Sigma_and_Angular_Dist}, the non-zero vacuum frequency is responsible for larger regions of the parameter space with $\kappa \neq 0$. The ELN angular distributions corresponding to the regions of the parameter space that  are unstable for $\omega \neq 0$ can be mapped into effective  ELN distributions that are unstable.}
\label{fig:Alpha_times_Sigma_w_neq_0_without_nu_x}
\end{figure*}

Figure~\ref{fig:Alpha_times_Sigma_w_neq_0_without_nu_x} highlights the role of the vacuum frequency in the development of flavor instabilities in the plane spanned by $\alpha_{\bar\nu_e}$ and $\sigma_{\bar\nu_e}$ for $\omega < 0$ ($\omega > 0$) on the left (right). One can see that the flavor instabilities appear around $D^z_0=0$ for $\omega<0$ and around $\zeta=0$ for $\omega>0$. It is conceivable that new unstable regions with respect to the ones displayed in the right panel of Fig.~\ref{fig:Alpha_times_Sigma_and_Angular_Dist} develop in the proximity of the loci with $D^z_0=0$ and $\zeta=0$ since these regions mark the transition from instability to stability in the limit $\omega \rightarrow0$.

\subsection{Neutrino flavor instabilities with non-electron flavors}
\label{sec:Slow_Instabilities_with_nu_x}
When non-electron flavors are taken into account, the development of flavor instabilities depends on $\alpha_{\nu_x}$ and $\sigma_{\nu_x}$. For the sake of simplicity, we fix $\alpha_{\nu_x}$ and $\sigma_{\nu_x}$ as indicated in Table~\ref{tab:Benchmark}, instead of letting them vary as we do for the parameters characteristic of the distribution of $\bar\nu_e$. Note that the inclusion of the non-electron flavors effectively changes the neutrino self-interaction strength; however, this has no effect on the linear growth rate. Since in this paper we only focus on the linear regime, for the sake of simplicity we adopt the same constant $\mu$ for both systems with and without non-electron flavors.

\begin{figure*}[b!]
\includegraphics[width=0.48\textwidth]{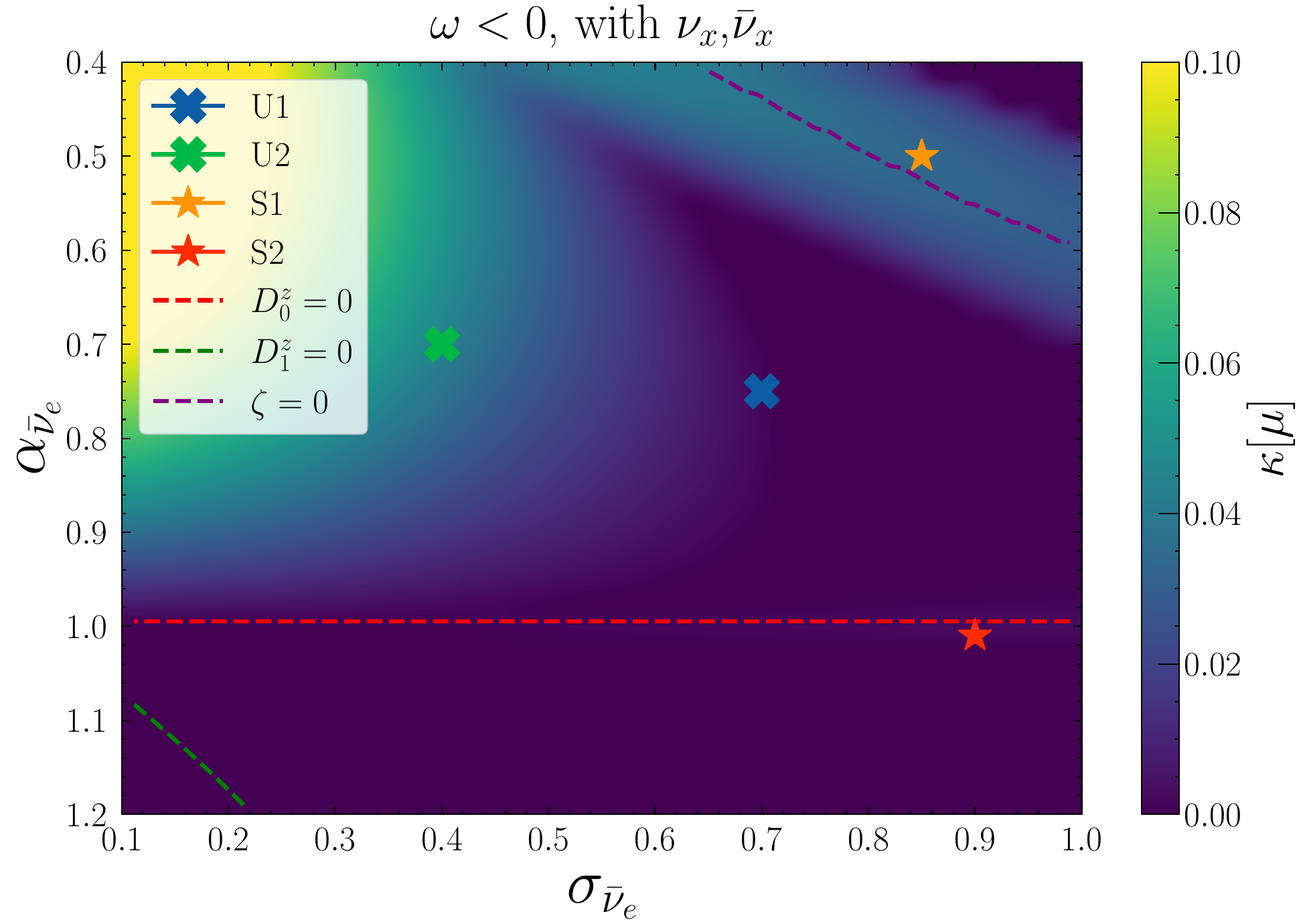}
\includegraphics[width=0.48\textwidth]{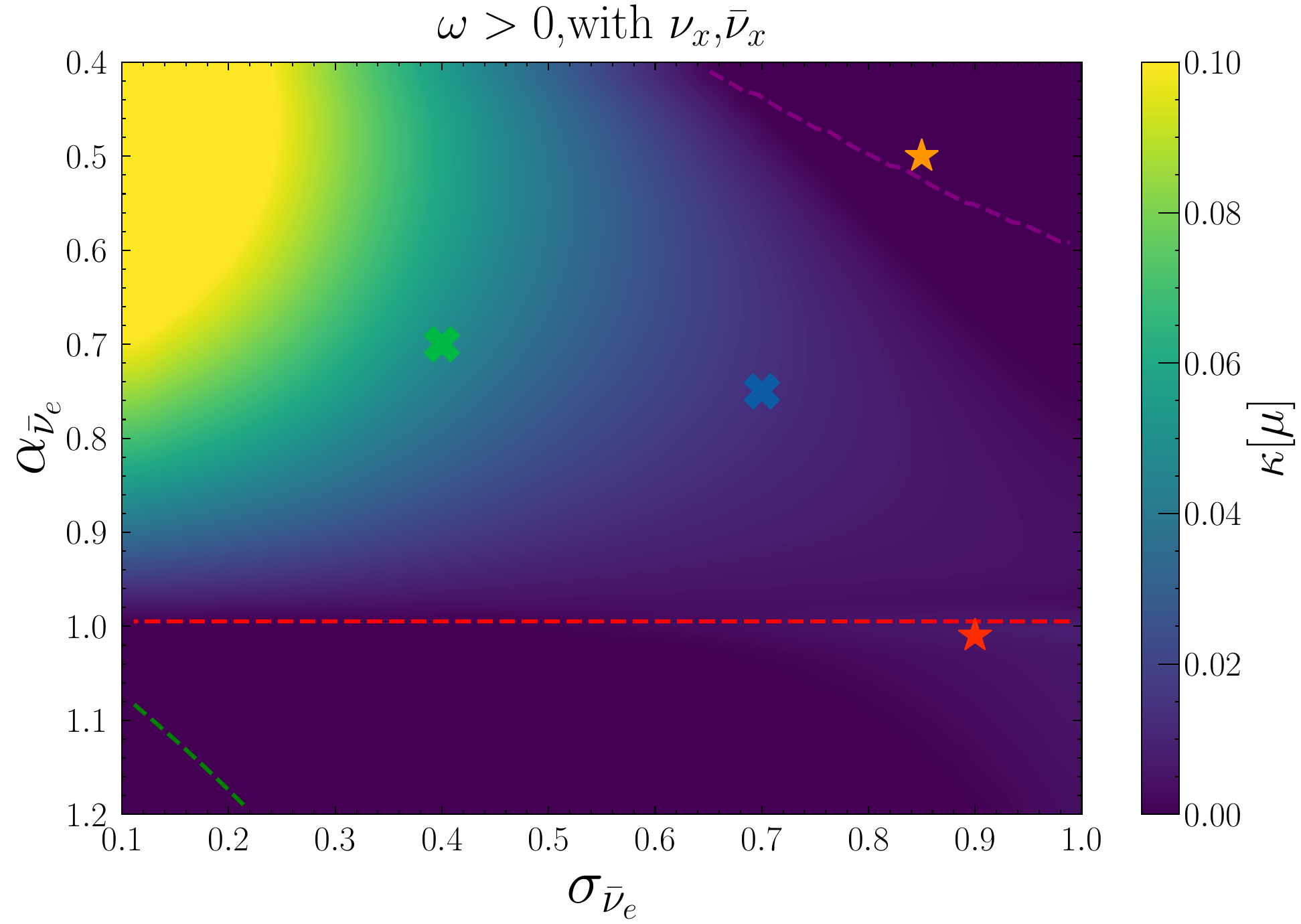}
\caption{Same as Fig.~\ref{fig:Alpha_times_Sigma_w_neq_0_without_nu_x} for the neutrino gas with $\nu_x$ and $\bar\nu_x$. Although both scenarios (with and without $\nu_x$ and $\bar\nu_x$) present the same instability pattern in the limit of $\omega=0$, the inclusion of $\omega\neq0$ results in different outcomes. For comparison the right panel of Fig.~\ref{fig:Alpha_times_Sigma_and_Angular_Dist} shows the instability growth rate for $\omega \rightarrow 0$.}
\label{fig:Alpha_times_Sigma_w_neq_0_with_nu_x}
\end{figure*}

The role of non-electron flavors in the development of flavor instabilities is visible in Fig.~\ref{fig:Alpha_times_Sigma_w_neq_0_with_nu_x} (cf.~also Figs.~\ref{fig:Alpha_times_Sigma_and_Angular_Dist} and \ref{fig:Alpha_times_Sigma_w_neq_0_without_nu_x}), in which one can see that the emergence of instabilities is opposite with respect to the case without $\nu_x$ and $\bar\nu_x$. For $\omega<0$ (left bottom panel), the instabilities around $D^z_0=0$ (cf.~left top panel) disappear, and new ones develop around $\zeta=0$. As for $\omega>0$ (right bottom panel), the instabilities around $\zeta=0$ disappear (cf.~right top panel). This behavior is in agreement with the dependence on EPN-XPN, as in the first-order correction in Eq.~\ref{eq:I_f_theta_first_order}, in which non-electron (anti)neutrinos cancel out the contribution from electron (anti)neutrinos. 

Recent work has already highlighted the importance of considering the angular distributions of non-electron (anti)neutrinos \cite{Capozzi:2020kge, Zaizen:2021wwl, Chakraborty:2019wxe, Zaizen:2022cik,Shalgar:2021wlj}. In Refs.~\cite{Capozzi:2020kge, Zaizen:2021wwl, Chakraborty:2019wxe}, the authors relax the assumption of equal $\nu_\mu$ and $\nu_\tau$ emission, showing that the onset of flavor instabilities depends on the combination of each initial flavor lepton number difference $\alpha$LN-$\beta$LN. Reference~\cite{Zaizen:2022cik} pointed out that the ELN-XLN angular distribution would be relevant even in the two-flavor approximation, considering a time-dependent ELN-XLN distribution to determine when the system achieves a steady state. Here, we show that the angular distributions of $\nu_x$ and $\bar\nu_x$ become important for the onset of flavor conversion triggered by the presence of crossings in the ELN-XLN angular distribution when vacuum mixing is taken into account, in agreement with the findings of Ref.~\cite{Shalgar:2021wlj}. Additionally, because of the inclusion of $\omega$, we show that the onset of flavor instabilities depends on the FPN and not only on the FLN.

\subsection{Effective neutrino angular distributions}
\label{sec:Effective_Angular_Dist}
We now discuss how the presence of a non-zero vacuum mixing term effectively has the role of reshaping the effective ELN angular distribution. This implies that stable configurations for $\omega =0$ on the verge of developing an ELN zero-crossing may develop an effective one because $\omega \neq 0$.

Equation~\ref{eq:I_f_theta_expansion} has the same structure as Eq.~\ref{eq:I_f_theta_w_0}, except for a correction factor. As a consequence, we can map the system with $\omega\neq0$ to an equivalent one with $\omega=0$ absorbing the correction factor in the definition of the initial angular distributions:
 \begin{equation}
 \label{eq:Effective_Angular_Dist}
 \rho^{0,\text{eff}}_{\alpha\alpha} = \rho^{0}_{\alpha\alpha}\sum_{k=0}^{\infty} \left(\frac{-\omega^c}{\Omega'+ vD^z_1}\right)^{k} = \rho^{0}_{\alpha\alpha}\left(1+\frac{\omega^c}{\Omega'+ vD^z_1}\right)^{-1}\ ,
 \end{equation} 
 \begin{equation}
 \label{eq:Effective_Angular_Dist1}
 \bar\rho^{0,\text{eff}}_{\alpha\alpha} = \bar\rho^{0}_{\alpha\alpha}\sum_{k=0}^{\infty} \left(\frac{\omega^c}{\Omega'+ vD^z_1}\right)^{k} = \bar\rho^{0}_{\alpha\alpha}\left(1-\frac{\omega^c}{\Omega'+ vD^z_1}\right)^{-1}\ .
 \end{equation} 
Relying on these definitions, Eq.~\ref{eq:I_f_theta_expansion} can be expressed in terms of  effective flavor difference spectra $g^{\text{eff}}$ and $\bar g^{\text{eff}}_v$:
\begin{equation}
 I_n(\Omega') = \mu \int \mathrm{d} v v^n \frac{\bar{g}^{\text{eff}}_v-g^{\text{eff}}_v}{\Omega'+\mu v D^z_1}\ .
\end{equation} 
The equation above suggests that, for $\omega \neq 0$, we can consider an effective ELN angular distribution, which depends on $\omega$ and which could be flavor unstable because of the vacuum term.

\begin{figure*}
\includegraphics[width=\textwidth]{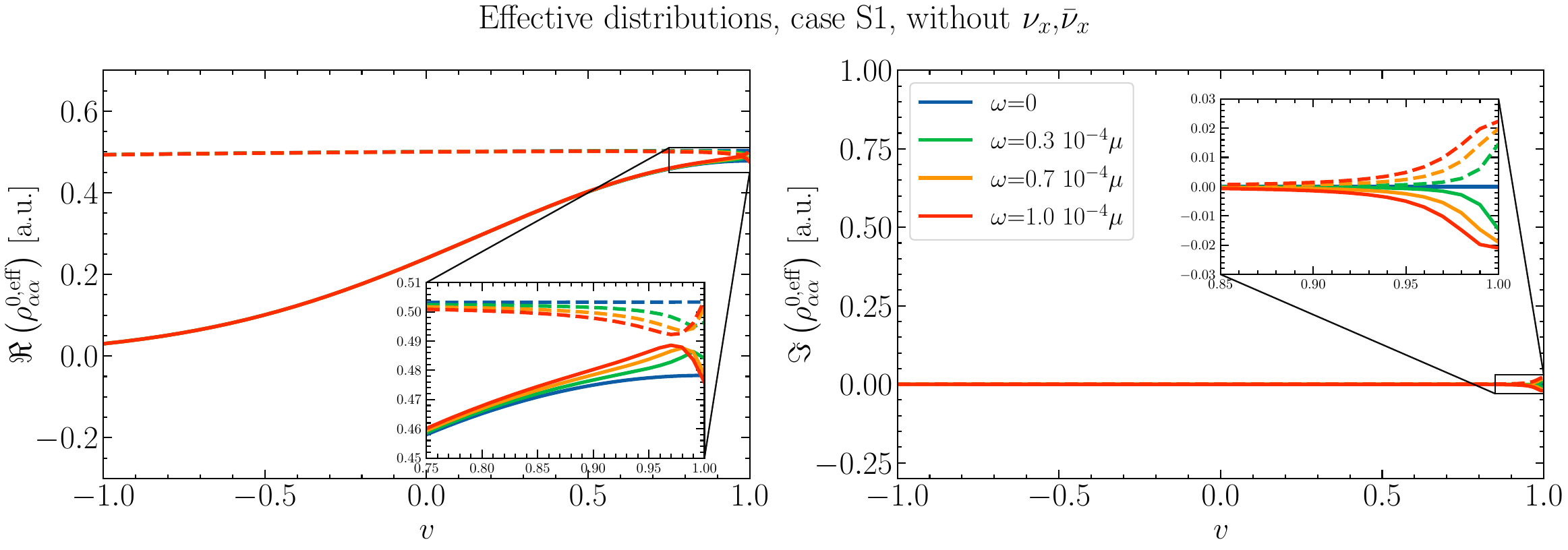}
\caption{Real (left panel) and imaginary (right panel) parts of the effective angular distributions of $\nu_e$ (dashed) and $\bar\nu_e$ (solid) for the configuration S1 without non-electron flavors. The different colors represent different values of $\omega$ between $0$ and $10^{-4} \mu$. The real component almost develops an ELN zero-crossing as $\omega$ increases, while the imaginary one only develops a small deviation from zero.
}
\label{fig:Eff_Ang_Dist}
\end{figure*}

Figure~\ref{fig:Eff_Ang_Dist} shows an example of the imaginary and real parts of the effective ELN-XLN angular distribution for the configuration S1 We note that the effective distribution develops a feature closely resembling a zero-crossing in the real part. However, the configuration S2 does not show any trend towards the development of an effective change of sign in $D^z_0$ or $D^z_1$; in this case the flavor pendulum is in a stable position, pointing downwards, and the perturbation induced by the vacuum term is not large enough to induce a flavor instability; see Sec.~\ref{sec:Vanishing_w}. 

In order to employ this parametrization, we should be aware of its caveats. The first is the regime of validity of the assumptions adopted to write $Q_v$ and $\bar Q_v$ for the linear equations. For stable solutions ($\kappa=0$), the ansatz in Eq.~\ref{eq:Q_v_ansatz} can be divergent due to a vanishing term in the denominator~\cite{Banerjee:2011fj, Airen:2018nvp}, hindering the validity of this approach within the assumption of small $\rho_{ex}$~\footnote{In cases where flavor instabilities are present, $|Q_v|$ takes the form of a Lorentzian function, making the set of solutions provided by Eq.~\ref{eq:Q_v_ansatz} suitable to look for instabilities.}. This problem stems from the assumption of the collective regime in Eq.~\ref{eq:LI_Solutions}, where all angular modes were considered to evolve with the same $\Omega$; this does not hold when flavor mixing is sufficiently small, and the off-diagonal elements of $H_{\nu\nu}$ are negligible. In such scenarios, the angular modes oscillate, independently of each other, with frequency $\gamma_v$. If an unstable solution exists, given enough time, it will grow and make the off-diagonal components of $H_{\nu\nu}$ large enough to enter the collective regime.  In this case, the effective angular distributions in Eq.~\ref{eq:Effective_Angular_Dist} contain an imaginary component, which is not physical. It is worth noticing that we cannot apply the stability criteria to the effective angular distributions, since they were developed under the consideration of real and positive $\rho_{\alpha\alpha}$. 

\section{Outlook}
\label{sec:Conclusions}
  Flavor conversion of neutrinos, triggered by crossings in the flavor lepton number (FLN) angular distribution,   has been widely investigated in the limit of zero vacuum frequency (fast conversion). However, the impact of vacuum mixing in the development of flavor instabilities triggered by FLN crossings is poorly understood. In the light of recent work pointing towards a modification of the fast flavor conversion phenomenology for non-vanishing vacuum frequency~\cite{Chakraborty:2019wxe,Airen:2018nvp,Zaizen:2021wwl,Shalgar:2020xns,Shalgar:2021wlj},  we explore the relevance of vacuum mixing on the onset of flavor instabilities in the presence of FLN zero-crossings.

Considering a mono-energetic, homogeneous, and axially symmetric neutrino gas, we
rely on a perturbative approach and show that the odd powers in $\omega$ are linked to the angular distribution of the neutrino flavor particle number (FPN). Hence, when $\omega \neq 0$, the conversion dynamics does not depend on the neutrino flavor lepton number (FLN) only~\cite{Padilla-Gay:2021haz}, but also on the particle number analogously to the traditional slow flavor conversion~\cite{Raffelt:2007yz}. 

Scanning a range of angular distributions for the electron flavors, we investigate how the flavor instability regions are modified by $\omega \neq 0$.  We find that $\omega \neq 0$ is responsible for inducing flavor instabilities in regions of the parameter space that are on the verge of instability for $\omega \rightarrow 0$.
This implies that the distinction between fast and slow flavor evolution is not binary, but continuous. 
We have shown that it is possible to have a flavor instability in the absence of ELN zero-crossings for large neutrino number density due to the specific shape of the effective ELN angular distribution and the self-interaction strength.
Deriving effective equations of motion that are formally identical to those of a system with $\omega\rightarrow0$ (where $\omega$ appears as a correction factor), we interpret the appearance of flavor instabilities for $\omega \neq 0$ as being due to the presence of a zero-crossing induced in the effective angular distribution which is a function of $\omega$.

\acknowledgments

This project has received support from the Funda\c{c}\~ao de Amparo \`a Pesquisa do Estado de S\~ao Paulo (FAPESP, Project 
No.~2022/01568-0 and No.~2022/09421-8), the Danmarks Frie Forskningsfond (Project 
No.~8049-00038B), the European Union (ERC, ANET, Project No.~101087058), and the Deutsche Forschungsgemeinschaft through Sonderforschungbereich SFB 1258 ``Neutrinos and Dark Matter in Astro- and Particle Physics'' (NDM). 
Views and opinions expressed are those of the authors only and do not necessarily reflect those of the European Union or the European Research Council. Neither the European Union nor the granting authority can be held responsible for them.

\appendix

\section{Role of the matter potential in  the development of flavor instabilities}
\label{sec:matter}

As discussed in Sec.~\ref{sec:Evolution_Equation}, the matter potential $\lambda$ is responsible for shifting the real part of the eigenfrequency ($\gamma$), not affecting the instability given by the imaginary  $\kappa$.  This Appendix shows this behavior explicitly. 

\begin{figure*}
\centering
\includegraphics[width=0.49\textwidth]{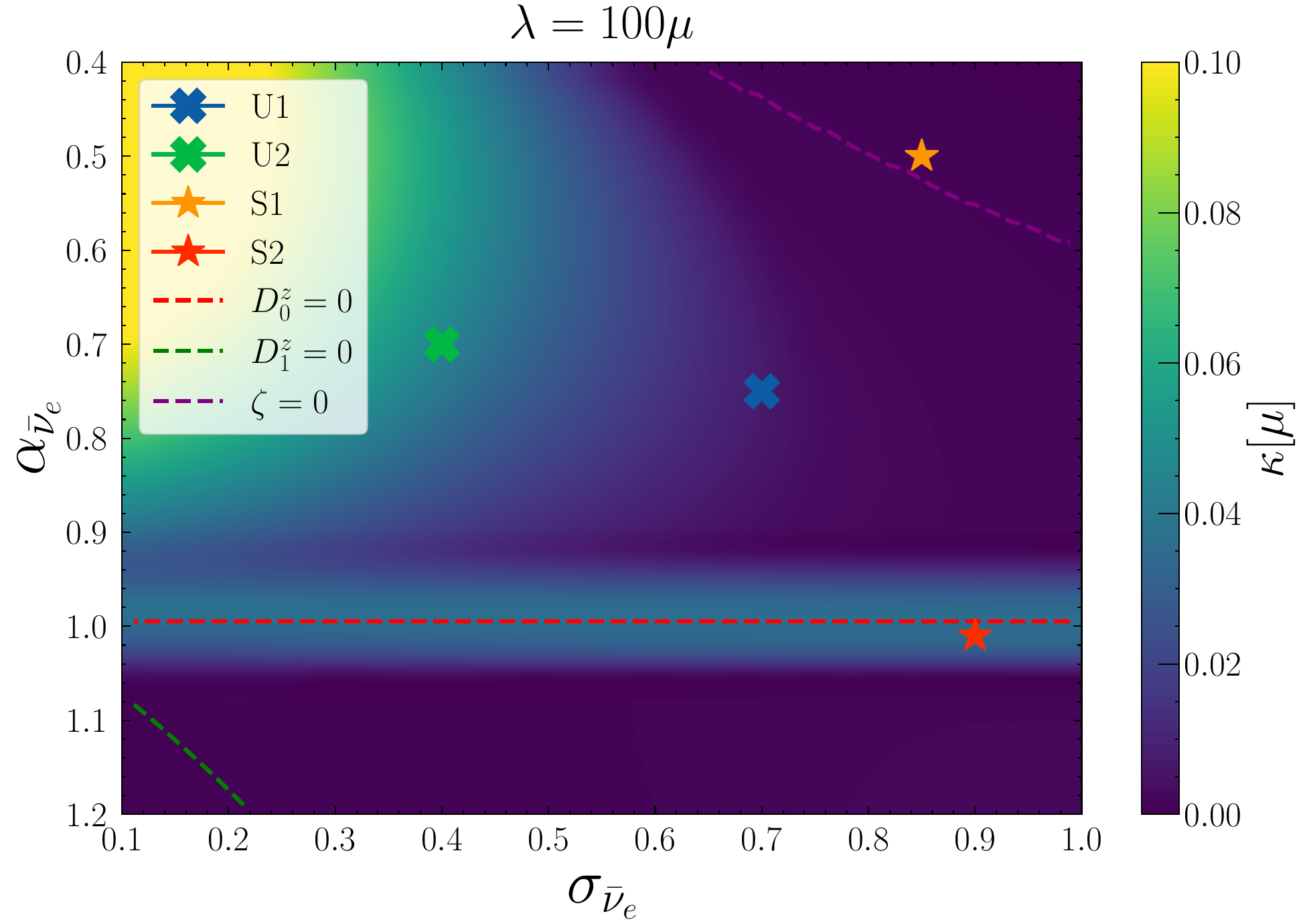}
\includegraphics[width=0.49\textwidth]{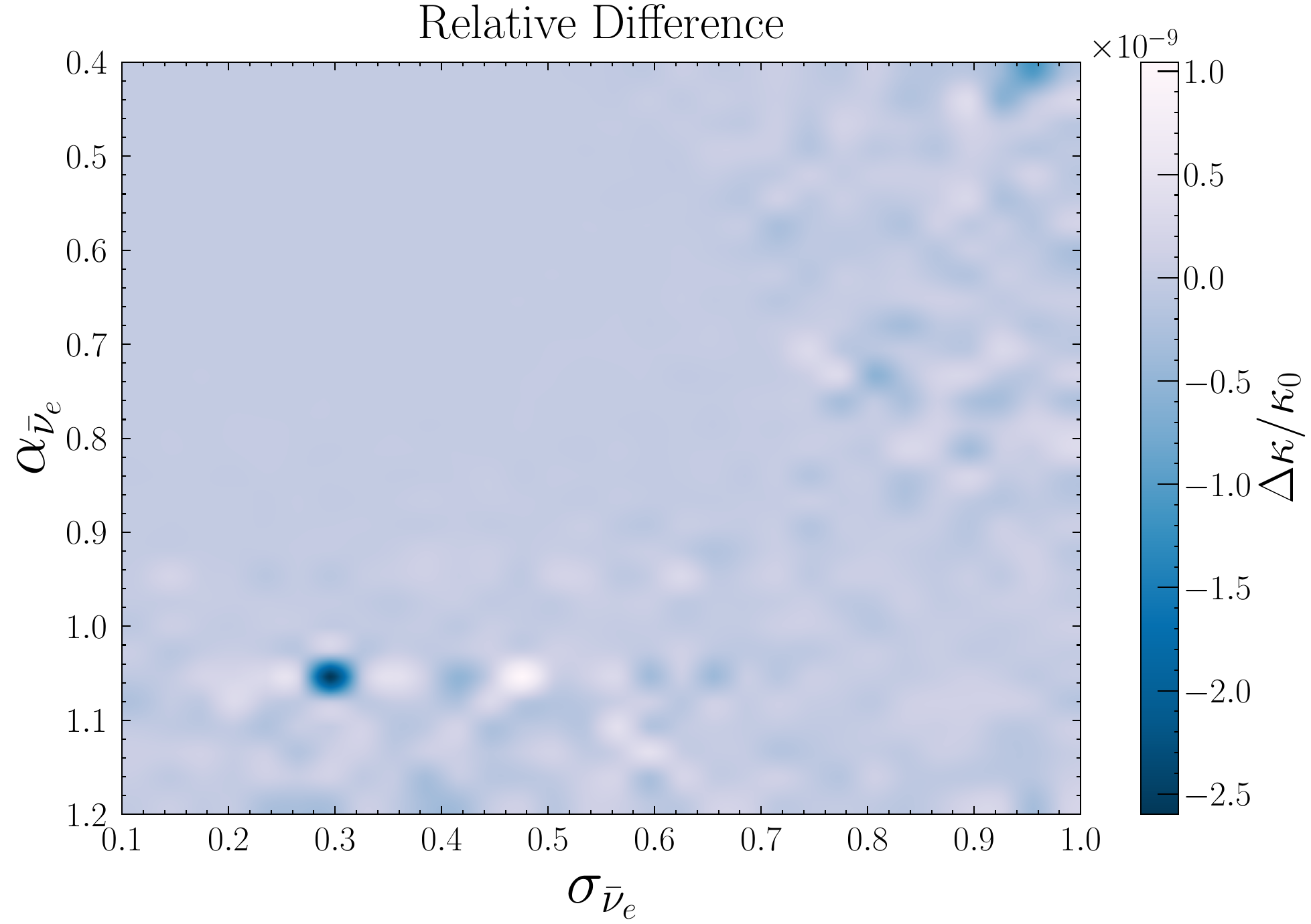}
\caption{{\it Left:} Same as Fig.~\ref{fig:Alpha_times_Sigma_w_neq_0_without_nu_x} for $\omega<0$ and  $\lambda=100\mu$. {\it Right:} Relative difference in the growth rate ($\Delta \kappa/\kappa_0$) for the cases with  ($\lambda=100 \mu$) and without ($\lambda=0$)  matter potential, taking the latter as reference. The matter potential does not affect the growth rate of flavor instabilities. }
\label{fig:Alpha_times_Sigma_lambda_neq_0}
\end{figure*}

In order to assess the impact of $\lambda$, we need to insert the latter in the denominator of Eq.~\ref{eq:I_f_theta}. The correspondent results for the growth rate are shown in Fig.~\ref{fig:Alpha_times_Sigma_lambda_neq_0}. For simplicity, we only consider the scenario with $\omega<0$ and without $\nu_x$ and $\bar\nu_x$. Nevertheless, the same trend is found in the other scenarios. The left panel of Fig.~\ref{fig:Alpha_times_Sigma_lambda_neq_0} shows that there is no appreciable difference in the growth rate between the cases with and without matter (cf.~Fig.~\ref{fig:Alpha_times_Sigma_w_neq_0_without_nu_x}). The right panel of Fig.~\ref{fig:Alpha_times_Sigma_lambda_neq_0} represents the relative difference of the growth rate for the cases with and without $\lambda$, taking $\lambda=0$ as our reference. As one can see, the relative difference is around $10^{-9}$, which is comparable to numerical precision.

\begin{figure*}
\centering
\includegraphics[width=0.7\textwidth]{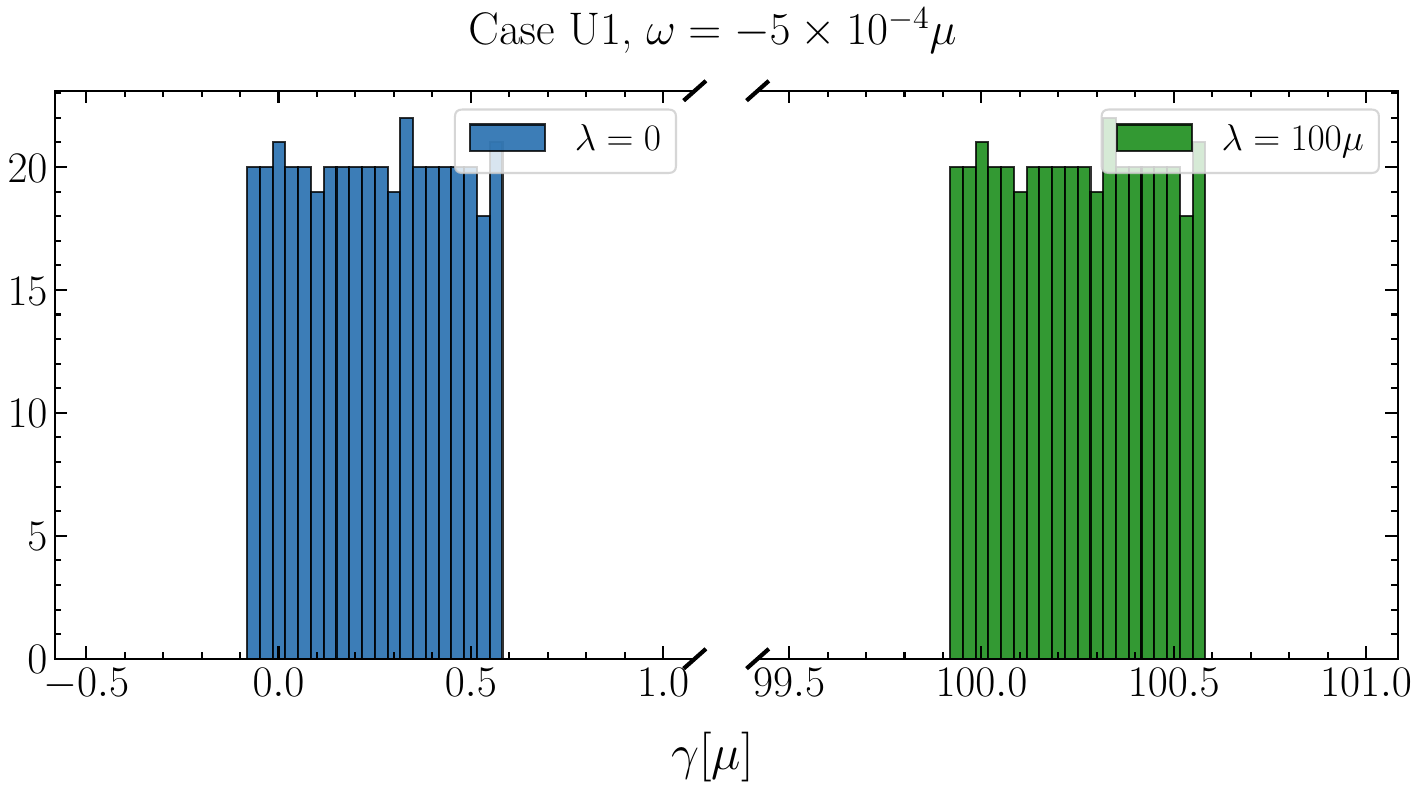}
\caption{Discrete spectrum of the eigenfrequencies $\gamma$ for the case U1 with $\omega = -5\times 10^{-4} \mu$, obtained via discretization of the linear equations of motion (Appendix~\ref{sec:Appendix_Numerical}) with $200$ angular bins. The effect of $\lambda$ is to shift $\gamma$.}
\label{fig:gamma_spectrum_U1}
\end{figure*}

Figure~\ref{fig:gamma_spectrum_U1} represents the discrete spectrum of the real part $\gamma$ of the eigenfrequencies obtained using $200$ angular bins. One can clearly see the shift $\gamma \rightarrow \gamma +\lambda$ when the matter potential is taken into account. This corroborates the conclusion that, in the linear regime, the matter potential does not affect the growth of flavor instabilities.

\section{Methods to compute the eigenfrequencies in the linear normal-mode analysis}
\label{sec:Appendix_Numerical}
In this appendix, we outline the approach adopted for determining the eigenfrequencies $\Omega$ in the linear stability analysis. First involving a semi-analytical integration of Eq.~\ref{eq:I_f_theta} to find the solution of Eq.~\ref{eq:Det_0}, and then relying on the discretization of the linear equations Eq.~\ref{eq:Linear_Equations}.

\subsection{Semi-analytical integration}
To solve Eq.~\ref{eq:Det_0}, it is necessary to evaluate Eq.~\ref{eq:I_f_theta} as a function of $\Omega$. One needs to solve this integral numerically, however, when numerical integration involves discretizing the integrand, it can lead to the emergence of spurious solutions due to issues related to the behavior of the discretized function near the branch-cut of the true continuous function~\cite{Morinaga:2018aug}.

To solve this problem, we use a semi-analytical approach with a piecewise constant approximation to $v^n g_v$. This approximation allows to factorize $v^n g_v$ out of the integral within each sub-interval, making it possible to solve the remaining part of the integral analytically, while preserving the discontinuity at the branch-cut. Following the approach outlined in Ref.~\cite{Morinaga:2018aug}, we divide the angular domain $[-1,1]$ into N sub-intervals ${[s_k,s_{k+1}]}$, centered on $v_k=(s_k+s_{k+1})/2$. This division enables us to express the components of the integrals in Eq.~\ref{eq:I_f_theta} as
\begin{widetext}
\begin{equation}
 \int_{-1}^{+1} \mathrm{d}v \frac{ u^n g_v}{v-x} \approx \sum_{k=1}^{N}v_k^n g_{v_k} \int_{s_k}^{s_{k+1}} \mathrm{d}v \frac{1}{v-x} = \sum_{k=1}^{N}v_k^n g_{v_k} \ln\left( \frac{s_{k+1}-x}{s_{k}-x}\right)\ .
\end{equation}
\end{widetext}
In order to ensure that $v^n g_v$ is approximately constant within each sub-interval, we employ $1000$ angular bins and impose that $\text{max}(|(v^n g_{v})_{k+1}-(v^n g_{v})_{k}|)<1$. Subsequently, we use these semi-analytical integrals to calculate the determinant of $D(\Omega)$ in Eq.~\ref{eq:Det_0}. To find the zero of this determinant, we utilize the Nelder-Mead minimization method, which is implemented in the \textit{minimize} routine of the \textit{scipy} library \cite{2020SciPy-NMeth}. However, it is important to note that the success of the minimization algorithm depends on having a good initial guess on $\Omega_0$. To obtain this initial guess, we adopt the method described in the following section.

\subsection{Discretization of the linearized equations of motion of neutrinos}
One way to find the eigenfrequency $\Omega$ is to solve the discretized version of Eq.~\ref{eq:Linear_Equations}. By discretizing these equations into $N$ angular bins, we obtain~\cite{Abbar:2015mca}:
\begin{equation}
\label{eq:Discretized_Linear_Equations}
 i\partial_t \rho^i_{ex} = (H^i_{ee}-H^i_{xx})\rho^i_{ex} - g_{v_i}\sum_{j=1}^{N}(1-v_iv_j)(\rho^j_{ex}-\bar\rho^j_{ex})\ ,
\end{equation}
\begin{equation}
\label{eq:Discretized_Linear_Equations1}
 i\partial_t \bar\rho^i_{ex} = (\bar H^i_{ee}- \bar H^i_{xx})\bar \rho^i_{ex} - g_{v_i} \sum_{j=1}^{N}(1-v_iv_j)(\rho^j_{ex}-\bar\rho^j_{ex})\ .
\end{equation}
Defining $\vec \rho_{ex} = (\rho^1_{ex},\bar\rho^1_{ex},...,\rho^N_{ex},\bar\rho^N_{ex})$ as a $2N$ dimensional vector containing the off-diagonal elements of neutrinos and antineutrinos in the angular basis, we can write 
\begin{equation}
 i \partial_t \vec \rho_{ex} = \Lambda \vec \rho_{ex}\ .
\end{equation}
The eigenvalues $\Omega_i$ can be found diagonalizing the $2N \times 2N$ matrix $\Lambda$ and using it as an initial guess for the method described in the previous section. To this purpose, we use $200$ angular bins, and for the matrix diagonalization, we employ the \textit{linalg} module provided by the \textit{scipy} library~\cite{2020SciPy-NMeth}.

As already pointed out in the literature, this method may also produce spurious solutions, and their relevance can become more pronounced when $\kappa$ is small. This is particularly concerning when the true solution approaches a branch-cut on the complex plane of $D(\Omega)$, as spurious instabilities tend to emerge~\cite{Morinaga:2018aug}. However, when $\kappa$ is sufficiently large, the true solution becomes more distinguishable from the spurious ones, providing a reliable initial guess. With this in mind, we establish a cut-off value $\kappa_{\text{cut}}=10^{-10}\mu$ for which we consider the solution to be valid. Even if non-spurious instabilities should exist with $\kappa<\kappa_{\text{cut}}$, they would occur at length scales of the order of $\mathcal{O}(10^6)$~km, which are irrelevant to our purposes.



\bibliography{biblio.bib}

\end{document}